\documentclass[twocolumn,showpacs,preprintnumbers,amsmath,amssymb]{revtex4}
\usepackage{graphicx}% Include figure files
\usepackage{dcolumn}% Align table columns on decimal point
\usepackage{bm}% bold math

\begin{document}
%\draft

%\onecolumn 

%\noindent

\title{Centrality dependence of charged-neutral particle fluctuations 
in 158$\cdot$A~GeV $^{208}$Pb\/+\/$^{208}$Pb collisions}

\author{ 
 M.M.~Aggarwal,$^{1}$
 Z.~Ahammed,$^{2}$
 A.L.S.~Angelis,$^{3}$
 V.~Antonenko,$^{4}$
 V.~Arefiev,$^{5}$
 V.~Astakhov,$^{5}$
 V.~Avdeitchikov,$^{5}$
 T.C.~Awes,$^{6}$
 P.V.K.S.~Baba,$^{7}$
 S.K.~Badyal,$^{7}$
% ...
} 

\medskip
\affiliation{(WA98 Collaboration)}
\medskip

\affiliation{$^{1}$~University of Panjab, Chandigarh 160014, India}
\affiliation{$^{2}$~Variable Energy Cyclotron Centre, Calcutta
   700064, India}
\affiliation{$^{3}$~University of Geneva, CH-1211 Geneva
   4,Switzerland}
\affiliation{$^{4}$~RRC ``Kurchatov Institute'',
   RU-123182 Moscow}
\affiliation{$^{5}$~Joint Institute for Nuclear Research,
   RU-141980 Dubna, Russia}
\affiliation{$^{6}$~Oak Ridge National
   Laboratory, Oak Ridge, Tennessee 37831-6372, USA}
\affiliation{$^{7}$~University of Jammu, Jammu 180001, India}
\affiliation{$^{8}$~University of M{\"u}nster, D-48149 M{\"u}nster,
   Germany}
\affiliation{$^{9}$~SUBATECH, Ecole des Mines, Nantes, France}
\affiliation{$^{10}$~University of Rajasthan, Jaipur 302004, Rajasthan,
   India}
\affiliation{$^{11}$~University of Lund, SE-221 00 Lund, Sweden}
\affiliation{$^{12}$~MIT Cambridge, MA 02139}
\affiliation{$^{13}$~Institute of Physics, Bhubaneswar 751005,
   India}
\affiliation{$^{14}$~University of Tsukuba, Ibaraki 305, Japan}
\affiliation{$^{15}$~Universiteit
   Utrecht/NIKHEF, NL-3508 TA Utrecht, The Netherlands}
\affiliation{$^{16}$~KVI, University of Groningen, NL-9747 AA Groningen,
   The Netherlands} 
\affiliation{$^{17}$~Gesellschaft f{\"u}r Schwerionenforschung (GSI),
   D-64220 Darmstadt, Germany}
\affiliation{$^{18}$~Nuclear Physics Institute, CZ-250 68 Rez, Czech Rep.}
\affiliation{$^{19}$~Institute for Nuclear Studies,
   00-681 Warsaw, Poland}
\affiliation{$^{20}$~University of Tennessee, Knoxville,
   Tennessee 37966, USA}

\date{\today}

\begin{abstract}

  Results on the study of localized fluctuations in the
  multiplicity of charged particles and photons produced in 
  158$\cdot A$~GeV/c Pb+Pb collisions are presented for varying
  centrality.  The charged versus neutral particle multiplicity 
  correlations in 
  common phase space regions of varying azimuthal size are analyzed 
  by two different methods. Various types of mixed events are 
  constructed to probe fluctuations arising from different sources. 
  The measured results are compared to those from simulations and  
  from mixed events. 
  The comparison indicates the presence of non-statistical
  fluctuations in both the charged particle and photon multiplicities in 
  limited azimuthal regions.
  However, no correlated charged-neutral fluctuations, a 
  possible signature of formation of disoriented chiral condensates,
  are observed. An upper limit on the production of disoriented chiral 
  condensates is set.

\end{abstract}

\maketitle

\section{INTRODUCTION}

  The large number of particles produced in relativistic heavy-ion 
  collisions at the Super Proton Synchrotron (SPS) and the Relativistic Heavy-Ion
  Collider (RHIC) provide an opportunity to  analyze and study, 
  on an event-by-event basis, fluctuations in physical observables, 
  such as particle multiplicities, transverse momenta,  and their
  correlations. These studies provide information on the dynamics
  of multi-particle production and may help to reveal the phase transition 
  from hadronic matter to quark-gluon plasma (QGP)~\cite{henning,stephanov}.
  The formation of hot and dense matter in high energy 
  heavy-ion collisions also has the possibility to create matter in a chiral 
  symmetry restored phase in the laboratory. After the initial stage of 
  the collision, the system cools and expands returning to the normal QCD vacuum 
  in which chiral symmetry is spontaneously broken. During this process, 
  a metastable state may be formed in which the chiral condensate is 
  disoriented from the true vacuum direction.
  This transient state would subsequently decay by emitting pions coherently within
  finite sub volumes or domains of the collision region. The possibility of formation
  of disoriented chiral condensate (DCC) has been discussed
  extensively in recent years
  \cite{anselm,bj,raj,huang1,raj2,gavin1,asakawa,koch,gavin-kapusta}.
  The detection and study of the DCC state would provide valuable
  information about the chiral phase transition and the vacuum structure of strong
  interactions.

  Theoretical studies~\cite{raj,huang1,raj2} suggest that isospin 
  fluctuations caused by formation of a DCC would produce clusters of coherent pions 
  in localized phase space domains. The formation of DCC domains 
  would be associated with large event-by-event fluctuations in the ratio 
  of neutral to charged pions. The probability distribution of the neutral pion 
  fraction, $f$, in such a DCC domain has been shown \cite{anselm} to follow 
  the relation:
\begin{equation}
P(f) =  \frac{1}{2\sqrt{f}} {\rm ~~~~~~~~~~where~~~~} f = N_{\pi^0}/N_{\pi}
\label{eqn1}
\end{equation}
  which is quite different from that of the normal pion production mechanism.
  For the normal pion production, where the production of $\pi^{0}$,$\pi^{+}$,
  $\pi^{-}$ are equally probable, the $f$ distribution is binomial peaking at
  $1/3$.

  In the experimental search for localized domains of
  DCC a practical approach is to search for events with large and localized
  fluctuations (localized in pseudo-rapidity ($\eta$) and azimuthal angle 
  ($\phi$) ) in the ratio of the number of photons to charged
  particles which would directly reflect fluctuations in the neutral to
  charged pion ratio. Typical event structures would be similar to the 
  Centauro and anti-Centauro events reported by the JACEE collaboration \cite{jacee}.
  Results from other cosmic ray experiments have not ruled out the
  possibility of the DCC formation mechanism \cite{augusto}.
  The accelerator based studies carried out in $p-\bar{p}$ \cite{minimax} and
  heavy ion \cite{WA98-3,NA49pt} reactions have investigated particle
  production over extended regions of phase space. These analysis were not
  sensitive to the presence of small domains of DCC localized in 
  phase space. A first search for evidence of localized domains 
  of DCC has been carried out at the SPS by the WA98 experiment in a
  detailed study of central Pb+Pb events ~\cite{WA98-12}. 
  The analysis showed the presence of localized non-statistical fluctuations
  in the multiplicity of both photons and charged particles. 
  However, the charged-neutral fluctuations were found not
  to be correlated event-by-event, as would be expected for a DCC production
  mechanism. An upper limit on the frequency of DCC formation in central Pb+Pb
  collisions was set. Recently there have been theoretical suggestions to look for DCC
  formation in events for intermediate centralities~\cite{notsocentral}.
  In this paper we present first results on the centrality dependence of 
  localized charged-neutral multiplicity fluctuations. It is  
  based on an analysis of event-by-event fluctuation in the
  relative number of charged particles and photons
  detected within the common acceptance of the photon and charged particle 
  multiplicity detectors of the WA98 experiment~\cite{wa98}.
 
  The paper is organized in the following manner: 
  In the next section we describe the detectors used
  for the present analysis, the centrality selection criteria, the data 
  reduction, and simulation. Section~III deals with the analysis 
  techniques, where two analysis methods are presented, one based on 
  the correlation of photons and charged particles, and the other based 
  on a discrete wavelet transformation analysis. In section~IV, we present in detail 
  the construction of mixed events used for this study. Section~V discusses 
  the ability of the mixed events to probe specific fluctuations.
  Final results and discussion are given in section~VI. A summary is 
  presented in section~VII.

\section{EXPERIMENTAL SETUP AND DATA REDUCTION}

    In the WA98 experiment at CERN~\cite{wa98}, the main emphasis
    has been on high precision, simultaneous detection of both hadrons 
    and photons. The experimental setup consisted of large acceptance 
    hadron and photon spectrometers, detectors for charged particle and 
    photon multiplicity measurements, and calorimeters for transverse and 
    forward energy measurements. The present study makes use of the data from 
    the photon multiplicity detector (PMD), the silicon pad multiplicity detector
    (SPMD) and the mid-rapidity calorimeter (MIRAC).

\subsection{Centrality Selection}

    The centrality of the interaction was determined from the total
    transverse energy ($E_{\mathrm T}$) measured by the mid-rapidity 
    calorimeter (MIRAC)~\cite{awes}.  The MIRAC measures both the transverse
    electromagnetic ($E_{\mathrm T}^{em}$) and hadronic 
    ($E_{\mathrm T}^{had}$) energies in the interval $3.5\le\eta\le  5.5$ 
    with a resolution of $17.9\%/\sqrt E$ and $46.1\%/ \sqrt E$,
    respectively, where $E$ is expressed in GeV. The centrality,  
    or impact parameter of the collision,
    has a strong correlation with the amount of $E_{\mathrm T}$ 
    produced. Events with large $E_{\mathrm T}$ 
    production correspond to the most central, small impact parameter,
    collisions~\cite{WA98-10}.

\begin{figure*}
\begin{center}
\includegraphics[scale=0.8]{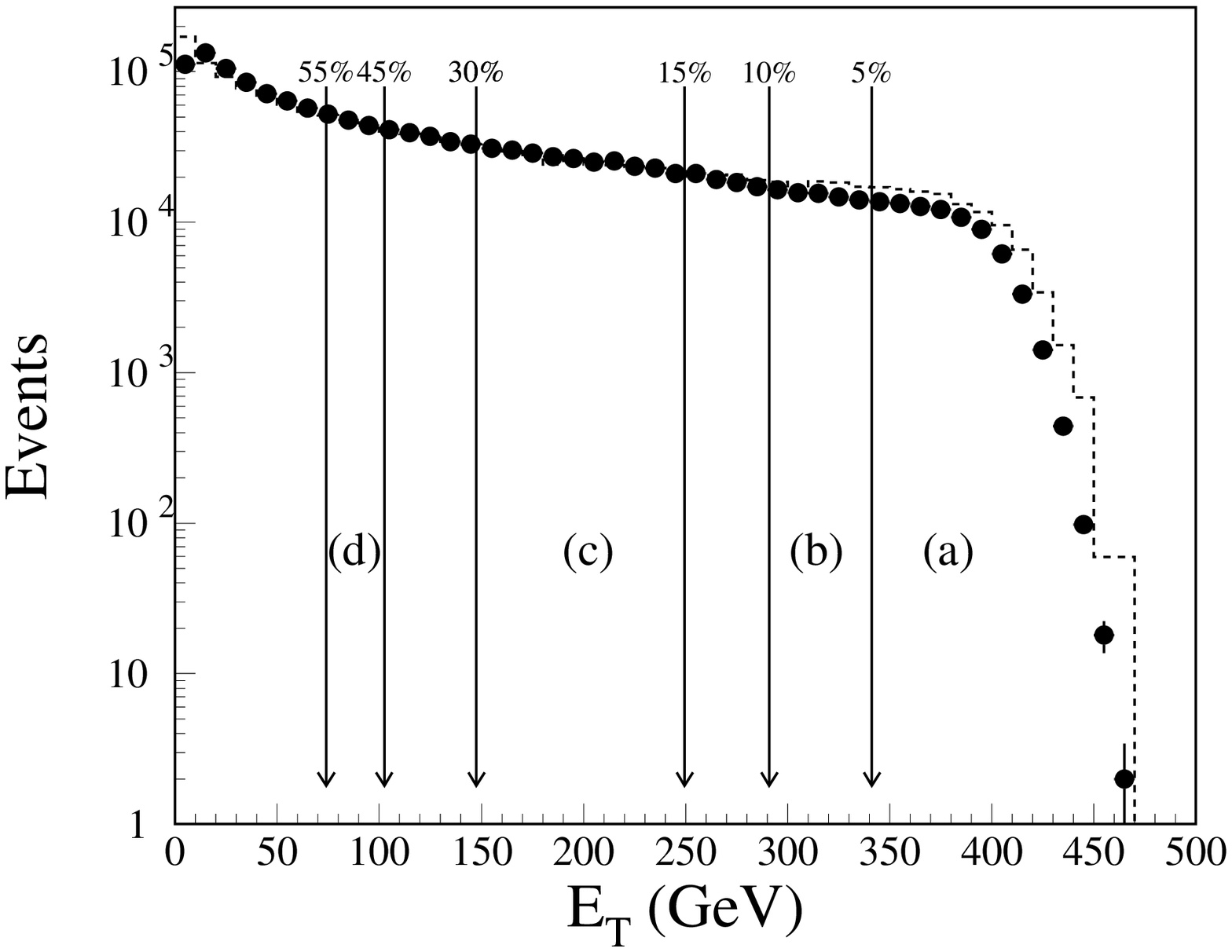}
\caption {
The total $E_T$ distribution (solid dots) measured in $3.5\le\eta\le 5.5$ 
for Pb+Pb collision at 158$\cdot$ A GeV/c. The total  $E_T$
distribution as obtained from VENUS is also shown as dashed
histogram. The $E_T$ values corresponding to the 
different centrality bins used in the analysis are shown.
(a) centrality-1 ($0-5\%$), (b) centrality-2 ($5-10\%$), (c) 
centrality-3 ($15-30\%$), (d) centrality-4 ($45-55\%$) of the minimum bias
cross section as determined by selection on the measured transverse energy
distribution. 
}
\label{et_cen}
\end{center}
\end{figure*}

    The centralities are expressed as fractions of the minimum bias cross section  
    as a function of the measured total $E_{\mathrm T}$.
    For the present analysis we have used data selections in four 
    centrality bins, the top $5\%$ (henceforth referred to as centrality-1),
    $5\%$ - $10\%$ (centrality-2), $15\%$ - $30\%$ (centrality-3), and 
    $45\% - 55\%$ (centrality-4) of the minimum bias cross section. 
    The minimum bias distribution of the total $E_{\mathrm T}$ is shown 
    in Fig.~\ref{et_cen}. The centrality bins used in this analysis 
    are marked in the figure.

\subsection{Photon Multiplicity Detector}

    The photon multiplicity is measured using the preshower photon 
    multiplicity detector (PMD) located at a distance of 21.5 meters 
    from the target. The PMD consists of 3 radiation lengths ($X_0$)
    thick lead converter plates in front of an array of square
    scintillator pads of four sizes, varying from 15 mm$\times$15 mm
    to 25 mm$\times$25 mm, placed in 28 box modules.
    Each box module consists of a matrix of $38\times$50 scintillator pads 
    read out using an image intensifier plus charged coupled device (CCD)
    camera system. The scintillation light is transmitted to the readout 
    device via a short wavelength shifting fiber spliced to a long extra-mural 
    absorber (EMA) coated clear fiber. The total light amplification
    of the readout system is $\sim$40000. Digitization of the CCD pixel
    charge is done by a set of custom built fast-bus modules employing an
    8 bit 20 MHz Flash ADC system. Details of the design and characteristics 
    of the PMD may be found in Ref. \cite{pmd_nim,WA98-9}. The results 
    presented here make use of the data from the central 22 box modules 
    covering the pseudo-rapidity range of  $2.9\le \eta\le 4.2$. The 
    clusters of hit pads, having total ADC content above a hadron rejection 
    threshold are identified as photon-like, the multiplicity being denoted 
    by $N_{\gamma-{\mathrm {like}}}$. 
    If the number of incident photons is denoted by $N_{\gamma}^{inc}$ and
    the number of photons detected above the hadron rejection threshold 
    as $N_{\gamma}^{th }$, then the photon counting efficiency 
    ($\epsilon_{\gamma}$) and purity of photon sample ($f_p$) are defined as,
    $\epsilon_{\gamma}~=~N_{\gamma}^{th}/N_{\gamma}^{inc}$ and 
    $f_p~=~N_{\gamma}^{th}/N_{\gamma-{\mathrm {like}}}$ respectively.
    These are estimated from detector simulations~\cite{pmd_nim,WA98-9}.
    The photon counting efficiencies 
    for the central to peripheral cases varies from $68\%$ to $73\%$. The 
    purity of the photon sample in the two cases varies from $65\%$ to $54\%$
    \cite{pmd_nim,WA98-9}.
    The acceptance in terms of transverse momentum ($p_{T}$) extends down to 
    about $30~MeV/c$, however the PMD energy resolution is not
    sufficient for particle-by-particle $p_{T}$ measurement.

\subsection{Silicon Pad Multiplicity Detector}

    The charged particle multiplicity ($N_{\mathrm {ch}}$) 
    is measured using the circular Silicon 
    Pad Multiplicity Detector (SPMD) located 32.8 cm from the target 
    and having full azimuthal coverage in the region $2.35< \eta < 3.75$, 
    corresponding to the central rapidity region at SPS energies 
    (where $\eta_{CMS} = 2.92$).
    The detector consists of four overlapping quadrants, each fabricated
    from a single 300~{$\mu m$} thick silicon wafer. The active area of each 
    quadrant is divided into 1012 pads forming 46 azimuthal wedges and 22 
    radial bins with a pad size increasing with radius to provide equal size
    pseudo-rapidity bins. The efficiency for detection of a charged particle 
    in the active area has been determined in test beam measurements to be better than 
    $99\%$.  Conversely, the detector is transparent to high energy photons, 
    since only about $0.2\%$ are expected to interact in the silicon. 
    During the data recording, $95\%$ of the pads worked properly and are 
    used in this analysis. Details of the characteristics of the SPMD can be 
    found in Ref. \cite{WA98-3,spmd_nim}. The SPMD
    does not provide $p_{T}$ measurement, but provides the multiplicity measurement 
    integrated over transverse momentum ($p_{T}$) with a threshold which 
    extends down to about $20~MeV/c$.

\subsection{Data Reduction}

\begin{figure*}
\begin{center}
\includegraphics[scale=0.4]{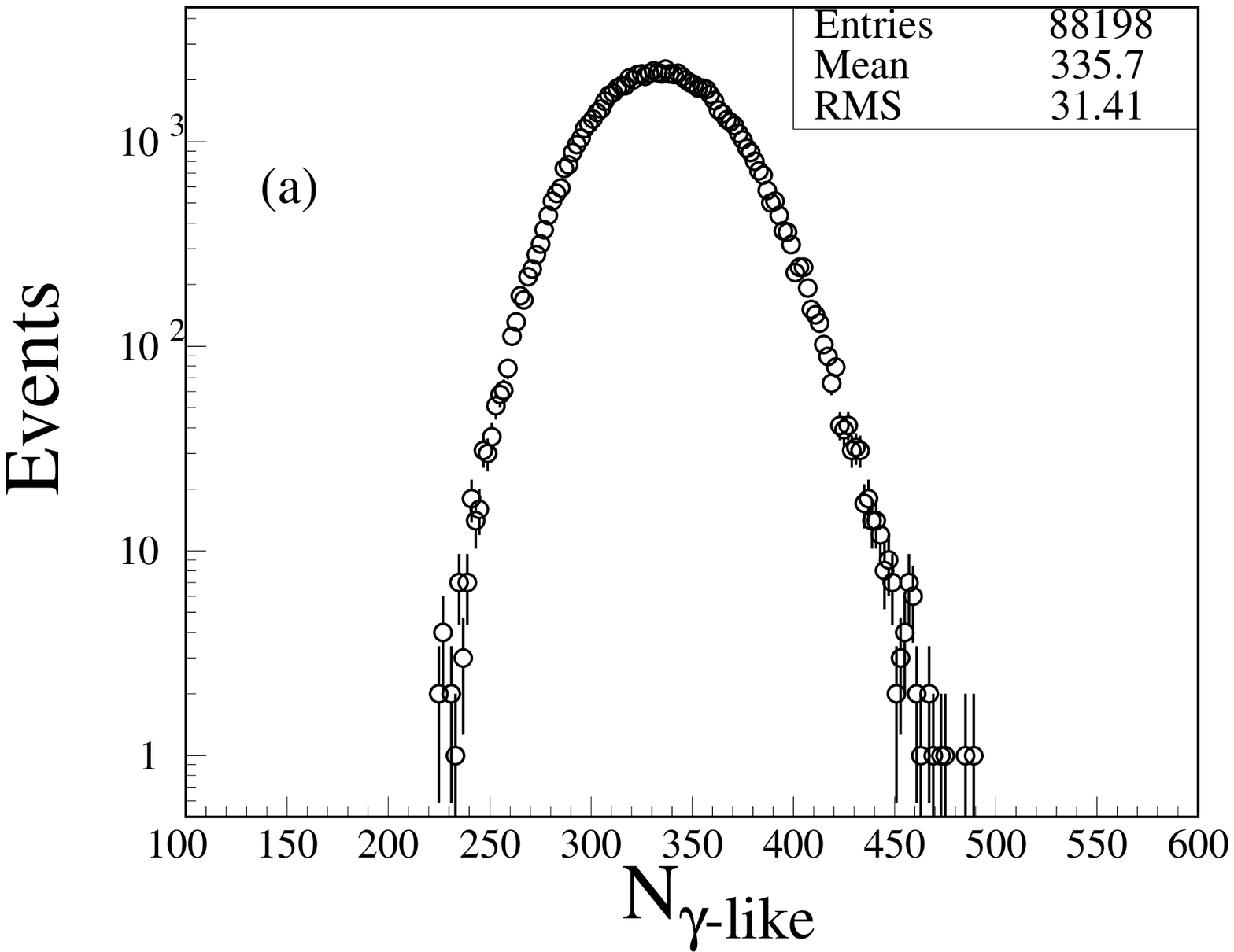}
\includegraphics[scale=0.4]{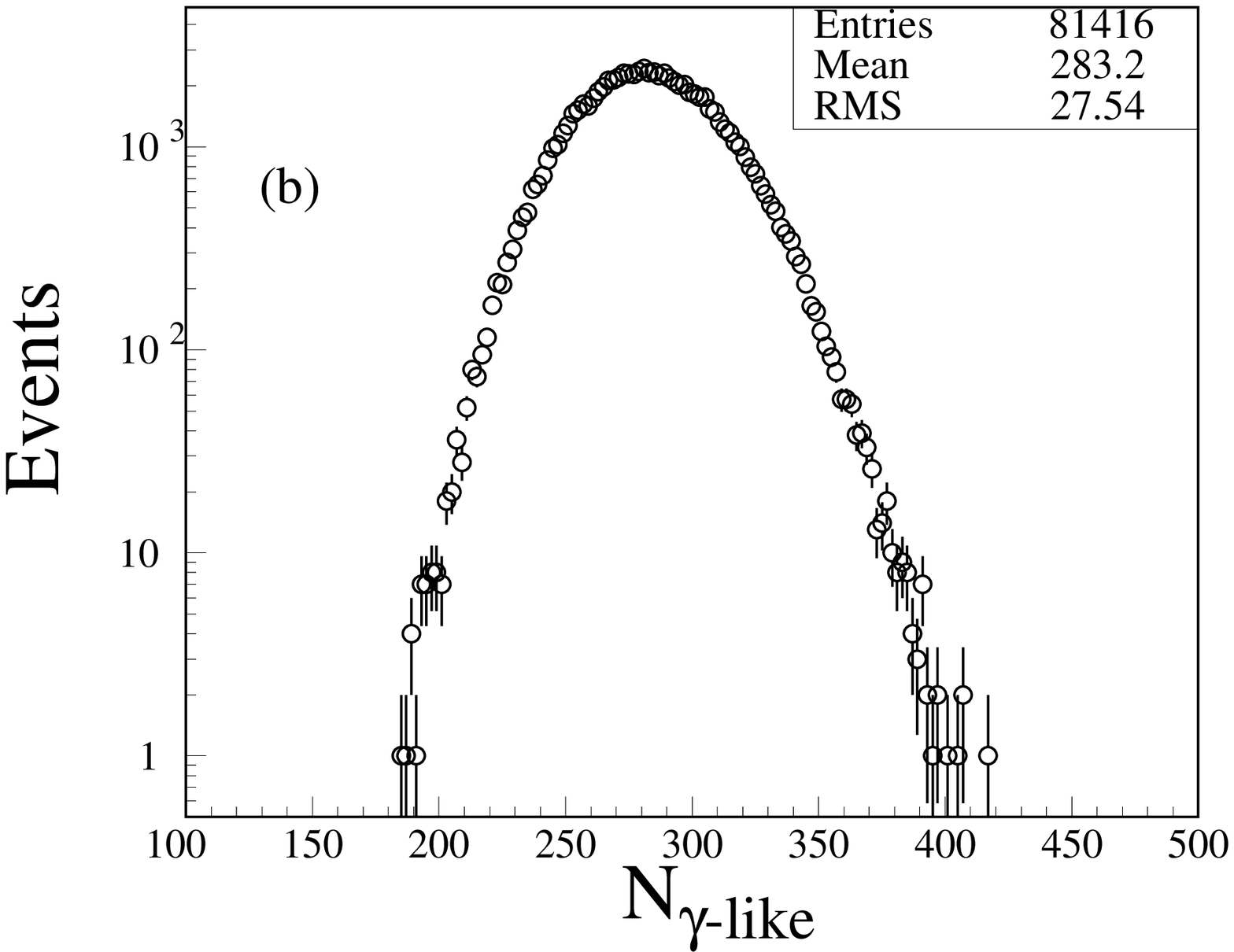}
\includegraphics[scale=0.4]{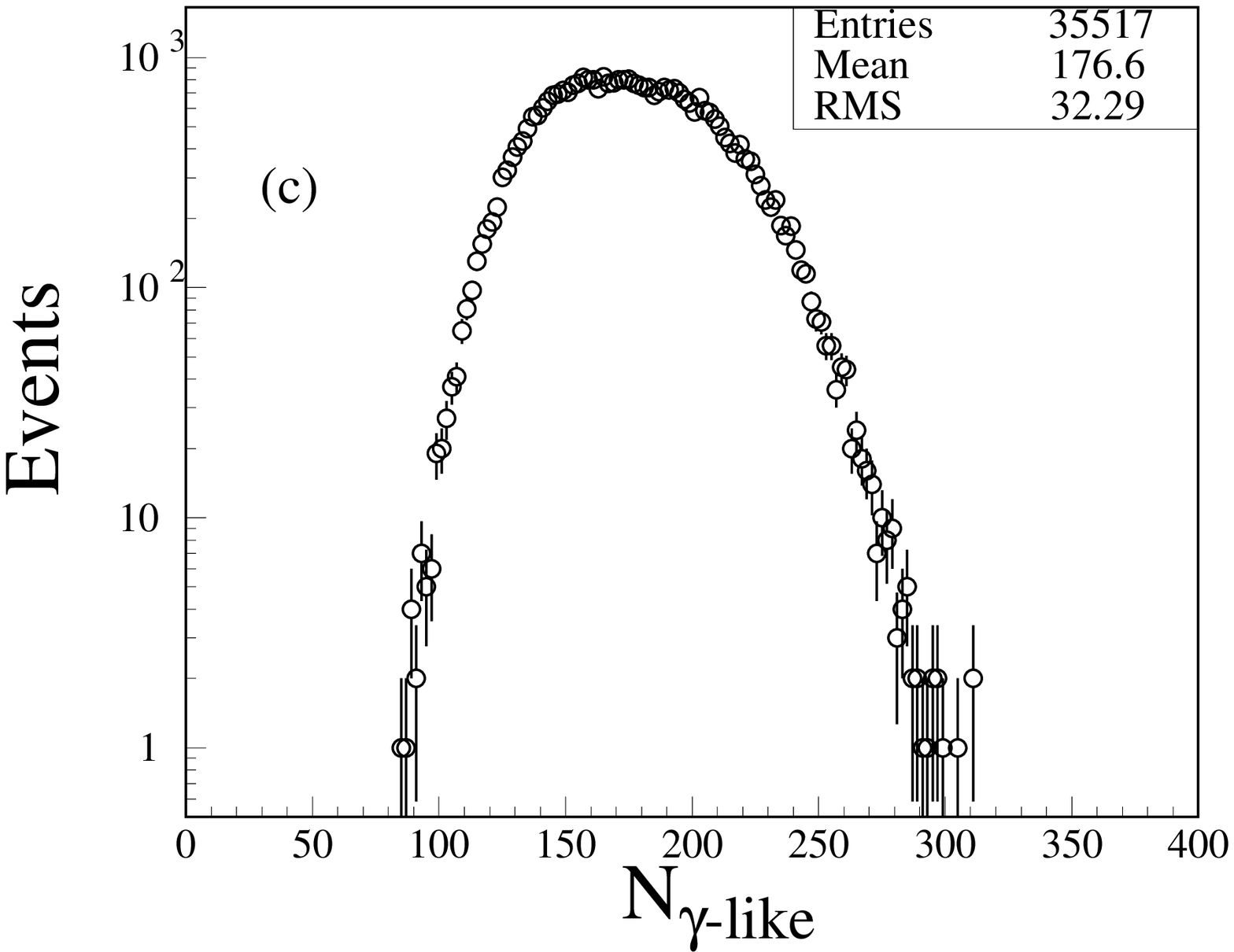}
\includegraphics[scale=0.4]{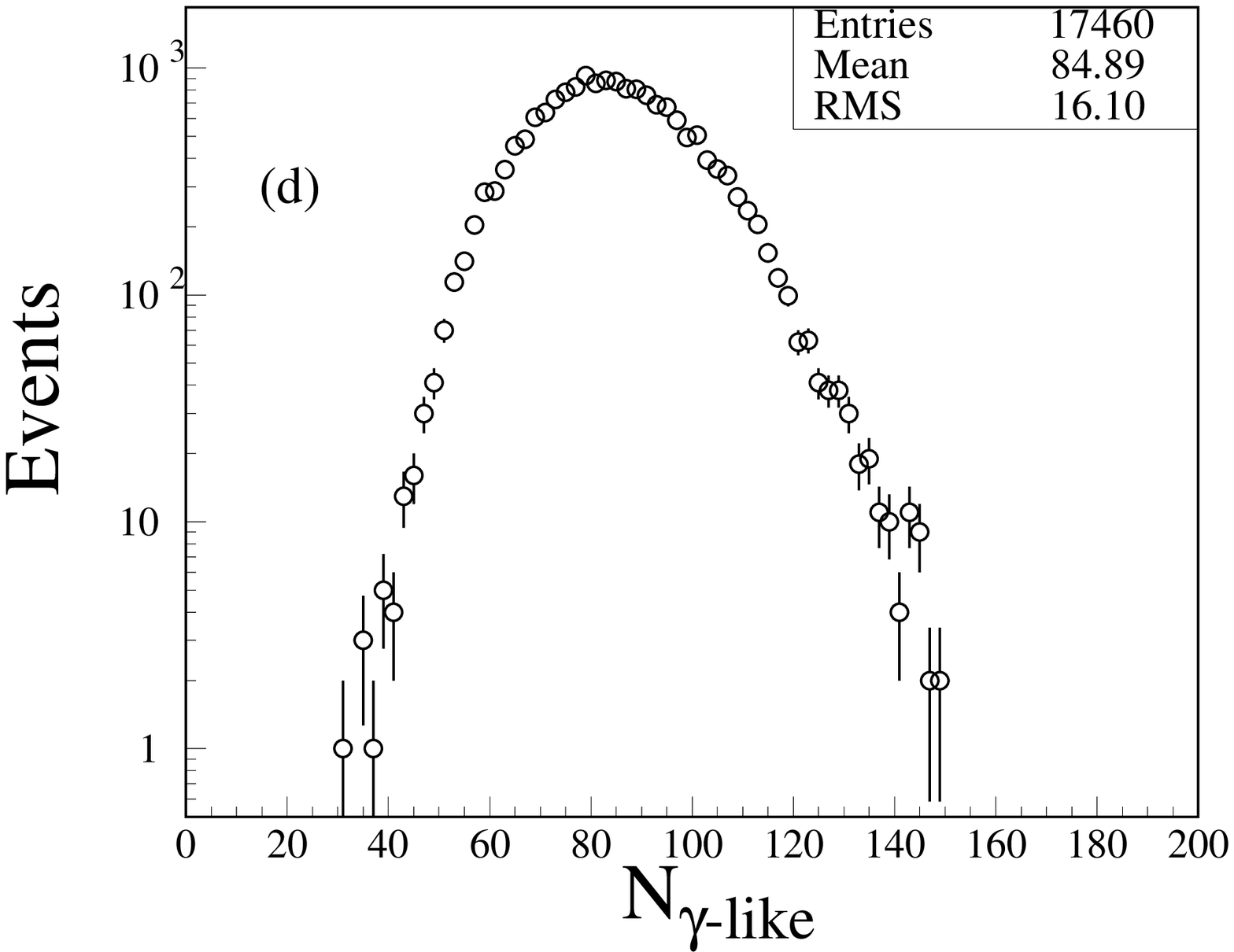}
\caption{The $N_{\gamma-{\mathrm {like}}}$ multiplicity distributions for
the four centrality selections for Pb+Pb collision at 158$\cdot$ A GeV/c.
(a) centrality-1 ($0-5\%$), (b) centrality-2 ($5-10\%$), (c) 
centrality-3 ($15-30\%$), (d) centrality-4 ($45-55\%$).
}
\label{ngam_cen}
\end{center}
\end{figure*}

\begin{figure*}
\begin{center}
\includegraphics[scale=0.4]{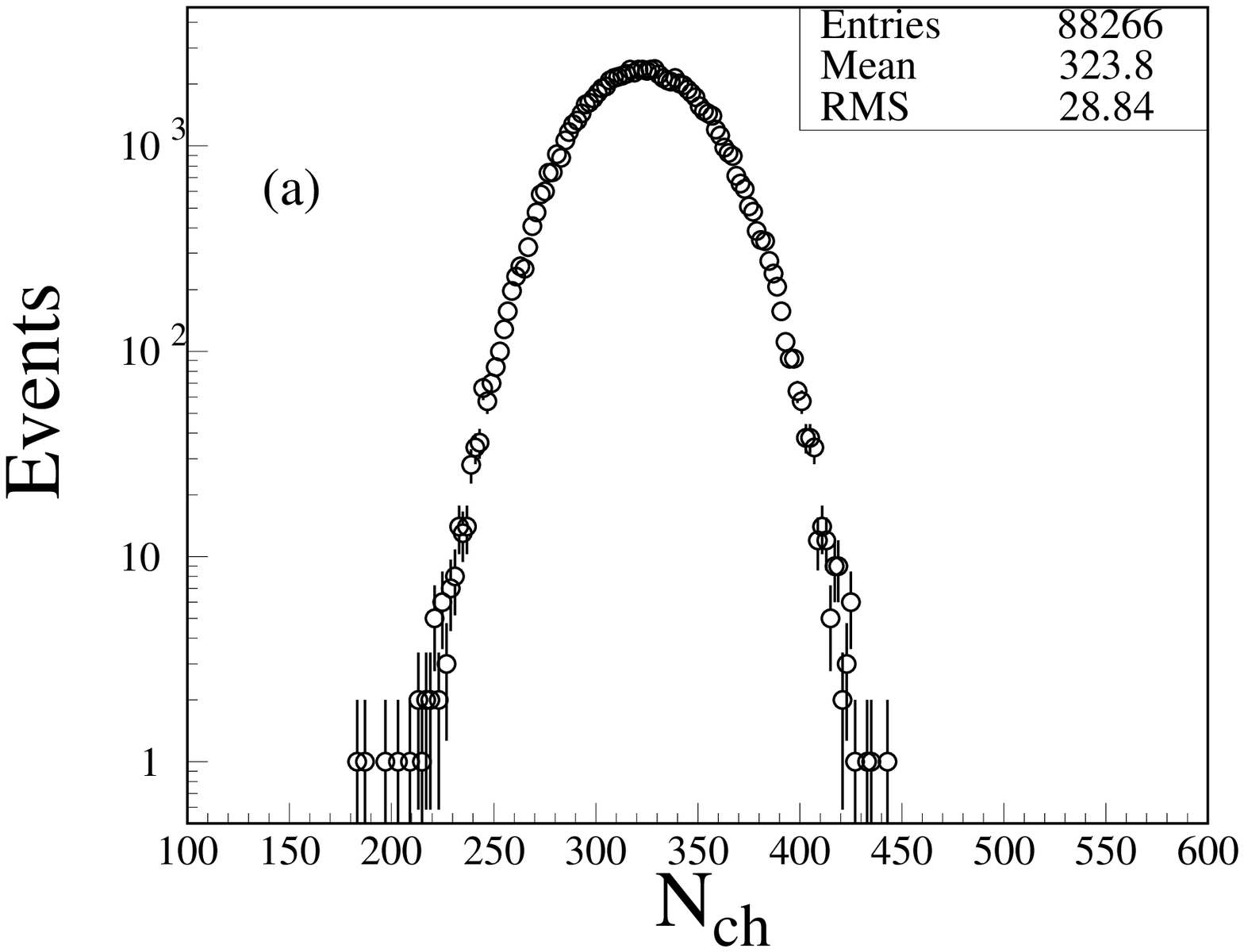}
\includegraphics[scale=0.4]{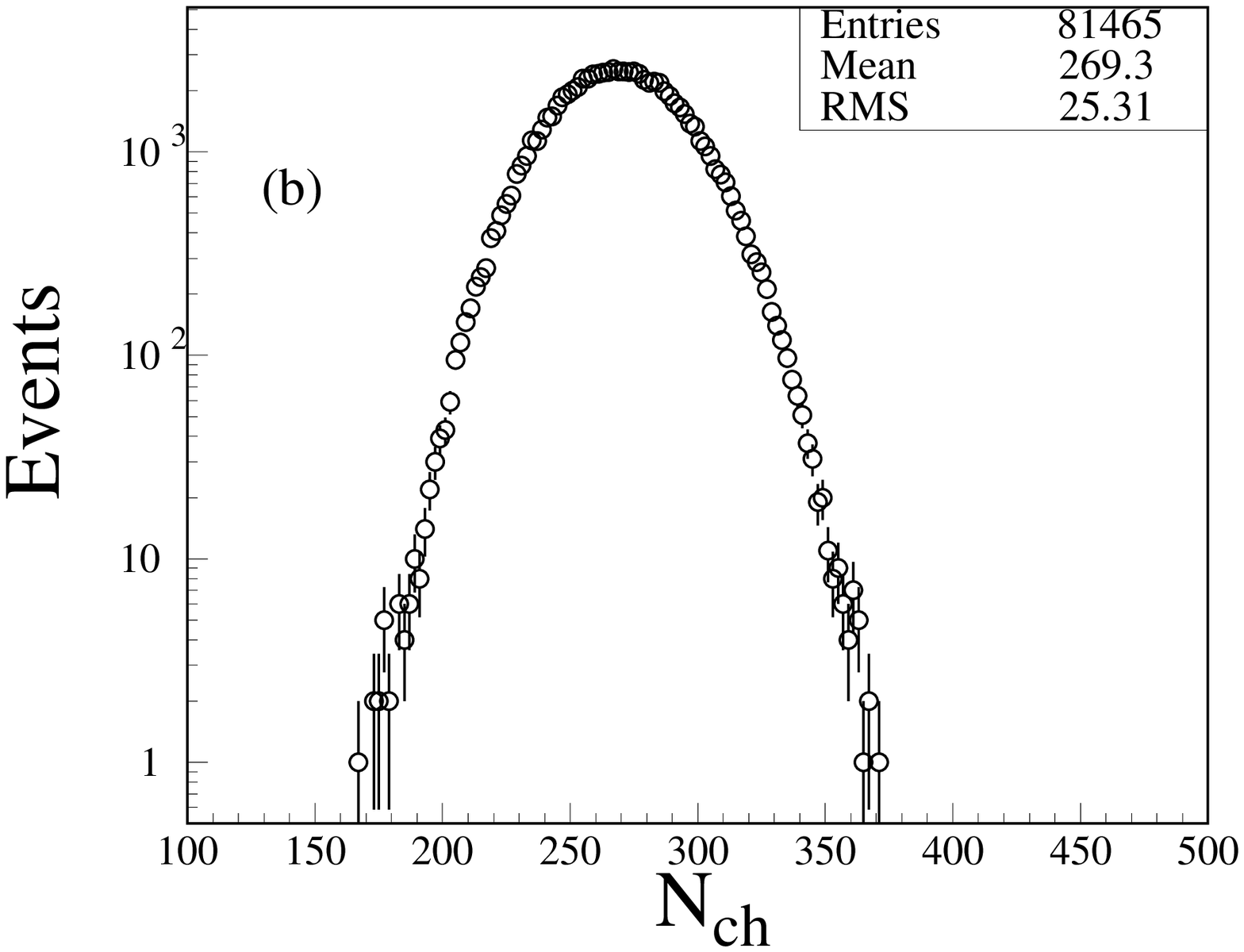}
\includegraphics[scale=0.4]{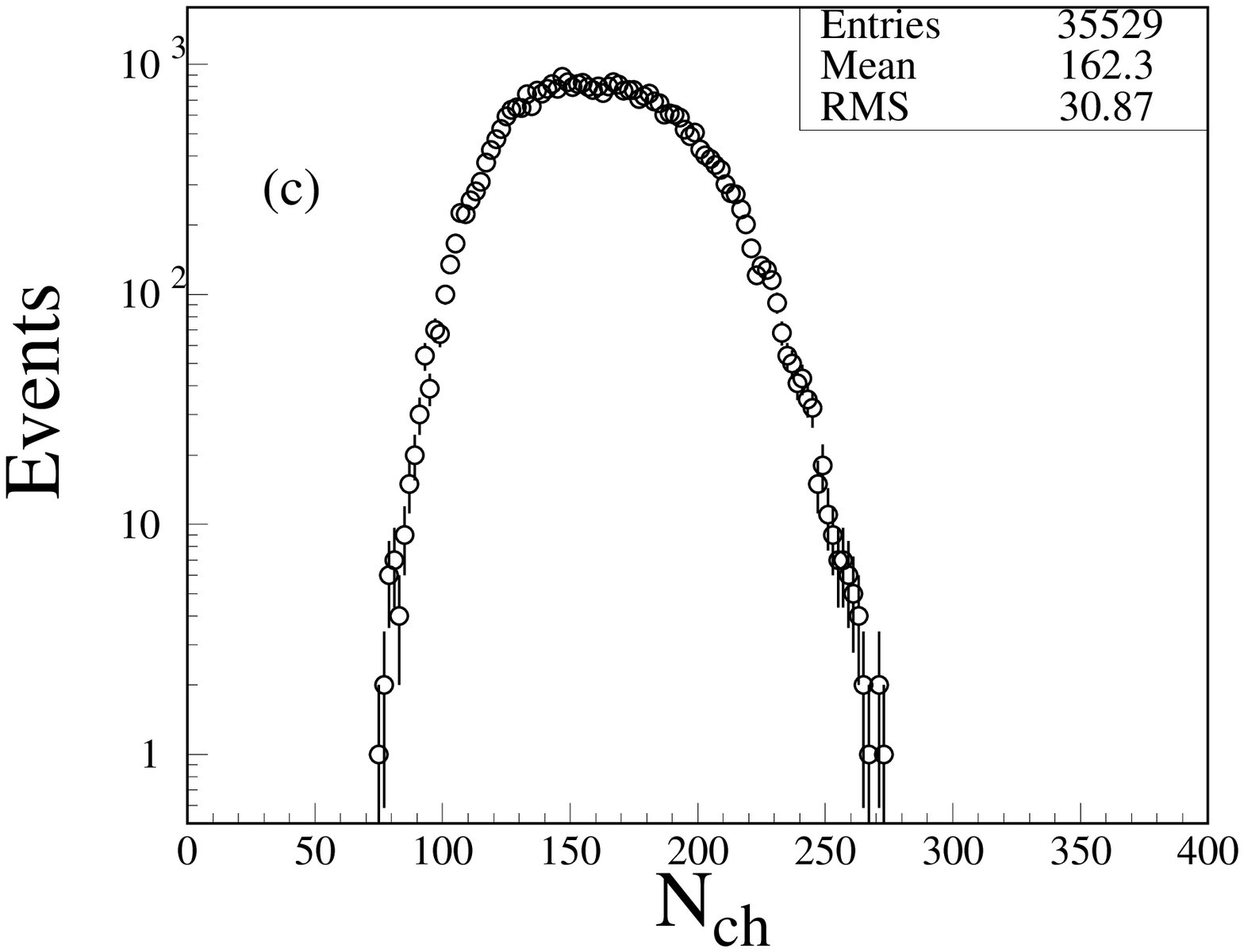}
\includegraphics[scale=0.4]{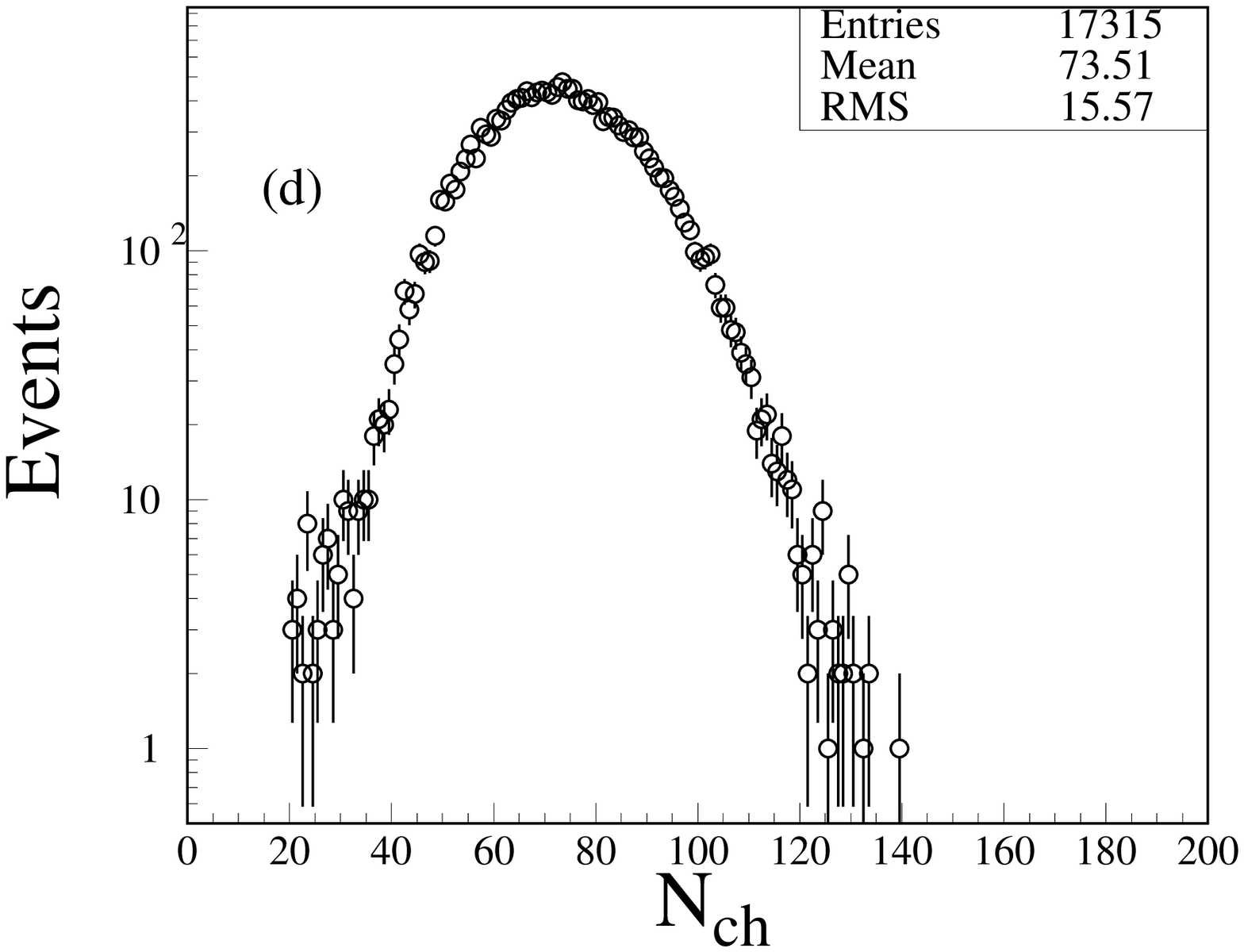}
\caption{The $N_{\mathrm {ch}}$ multiplicity distributions for the four 
centrality selections for Pb+Pb collision at 158$\cdot$ A GeV/c.
(a) centrality-1 ($0-5\%$), (b) centrality-2 ($5-10\%$), (c) 
centrality-3 ($15-30\%$), (d) centrality-4 ($45-55\%$).
}
\label{nch_cen}
\end{center}
\end{figure*}

    The data presented here were taken during December 1996 at the CERN
    SPS with the 158$\cdot A$~GeV Pb ion beam on a Pb target of thickness
    213~$\mu$m. The WA98 Goliath magnet was switched off during these runs.
    Events with beam pile-up, downstream interactions, and pile-up in the
    CCD camera system were rejected in the off-line analysis~\cite{WA98-3,WA98-9}. 
    The data have been analyzed for the region of
    common $\eta$ ($2.9< \eta < 3.75$) and $\phi$ coverage of the SPMD
    charged particle and PMD photon multiplicity detectors. 
    The $N_{\gamma-{\mathrm {like}}}$
    and the $N_{\mathrm {ch}}$ distributions for the four centrality
    bins are shown in Figs.~\ref{ngam_cen} and \ref{nch_cen}.
    The number of events analyzed, the mean number of photons and 
    charged particles along with the
    root mean square deviations are shown in the figures.
    The PMD and SPMD detectors provide momentum integrated multiplicity 
    measurements with very low thresholds. Since pions from DCC domains
    are expected to have small $p_T$ values, below the pion mass, the 
    momentum integration will dilute the DCC signal. On the other hand, 
    the large coverage of the PMD and SPMD are important to overcome the
    limitations of small number fluctuations.

    The various sources of systematic errors associated with the 
    $N_{\gamma-{\mathrm {like}}}$ and $N_{\mathrm {ch}}$ distributions
    have been investigated and described in detail previously~\cite{WA98-9,WA98-15}. 
    These include: 
\begin{itemize}

\item    (a) The uncertainty in the energy calibration and the associated 
        uncertainty in the energy threshold for hadron rejection
        in the PMD leads to an error in the efficiency for 
        $N_{\gamma-{\mathrm {like}}}$ clusters. The nominal hadron rejection
        threshold was set at three times the minimum ionizing
        particle (MIP) peak. The value of the threshold was reduced by 
        $10\%$~\cite{WA98-9} in order to 
        estimate the systematic error. The associated error in 
        $N_{\gamma-{\mathrm {like}}}$ is $2.5\%$.
\item    (b) The error due to the effect of clustering of pad signals in the PMD
        is a major source of error in $N_{\gamma-{\mathrm {like}}}$. This
        error is determined from GEANT~\cite{geant} 
        simulation by comparing the number 
        of known photon tracks on the PMD with the total number of
         reconstructed photon-like
        clusters. It is found that the number of clusters exceeds the 
        number of tracks by $3\%$ in the case of peripheral events and by
        $7\%$ for high multiplicity central events. 
\item   (c)  The error due to the variation in pad-to-pad gains of the scintillators
             in PMD was found to be less than $1\%$.
\item   (d)  The uncertainty in the  $N_{\mathrm {ch}}$ determination with the
             SPMD has been estimated to be about $4\%$  \cite{WA98-3}. 
\item   (e)  The error due to the
        finite resolution in the measurement of the total transverse 
        energy($E_T$) in MIRAC \cite{WA98-9} translates into an uncertainty
        in the centrality selection. The effect of this systematic error has 
        been determined by performing the analysis with varying  
        centrality cut within the MIRAC resolution. 
\end{itemize}
      The contribution of each of these various systematic errors to the final 
      results are discussed in the following sections. 

\subsection{Simulated Events}

    Simulated events were generated using the VENUS 4.12 event
    generator~\cite{venus} with the default parameter values. The output was processed
    through a detector simulation package in the GEANT 3.21 \cite{geant}
    framework. This simulation includes the full WA98 experimental setup and 
    includes experimental effects such as photon conversions, downstream 
    interactions, hadron backgrounds in the PMD, etc. which might dilute or 
    enhance the observed fluctuations.
    The effect of Landau fluctuations in the energy loss of charged 
    particles in  silicon  was included in the SPMD 
    simulation \cite{WA98-3}. For the PMD simulation, the GEANT results in 
    terms of energy deposition in pads were converted to the pad ADC 
    values using the MeV-ADC calibration relation. After this the ADC
    distribution is convoluted with a Gaussian function of proper width taken from the 
    readout resolution curve. If the energy deposition is less than 3 MIP, a
    Landau distribution is used for convolution. 
    The details of the PMD simulations
    taking into account the detector and readout effects can be found in 
    Ref.~\cite{pmd_nim}.
    The centrality selection with the simulated data has been made in an
    identical manner to the data, determined from the simulated total 
    transverse energy in MIRAC. The minimum bias total $E_{\mathrm T}$ 
    distribution  predicted by VENUS 
    is shown by the dashed histogram in
    Fig.~\ref{et_cen}. The agreement with the data is seen to be quite 
    reasonable. A total of 60K VENUS events with simulated detector response 
    were generated for the present study. These simulated events
    (henceforth referred to simply as VENUS events unless otherwise specified)
    were then processed with the same analysis codes as used for the analysis
    of the experimental data.

\section{ANALYSIS TECHNIQUES}

    Two different analysis methods have been used in the present study.
    In the first analysis method, the magnitude of the 
    $N_{\gamma-{\mathrm {like}}}$ versus $N_{\mathrm {ch}}$ multiplicity 
    fluctuations is obtained in decreasing phase space regions.  
    The second method employed the discrete wavelet transformation 
    technique to investigate the relative magnitude of the 
    $N_{\gamma-{\mathrm {like}}}$ versus $N_{\mathrm {ch}}$ fluctuations
    in adjacent phase space regions. The results from these methods
    of analysis applied to data, VENUS and various sets of 
    mixed events (to be discussed later) are compared to draw 
    proper conclusions.

\subsection{$N_\gamma$ versus $N_{\mathrm {ch}}$ correlations}

   In order to search for localized fluctuations
   in the photon and charged particle multiplicities, which may have
   non-statistical origin,  the correlation 
   between $N_{\gamma-{\mathrm {like}}}$ and $N_{\mathrm {ch}}$
   is investigated at various scales in $\phi$. 

\begin{figure*}
\begin{center}
\includegraphics[scale=0.5]{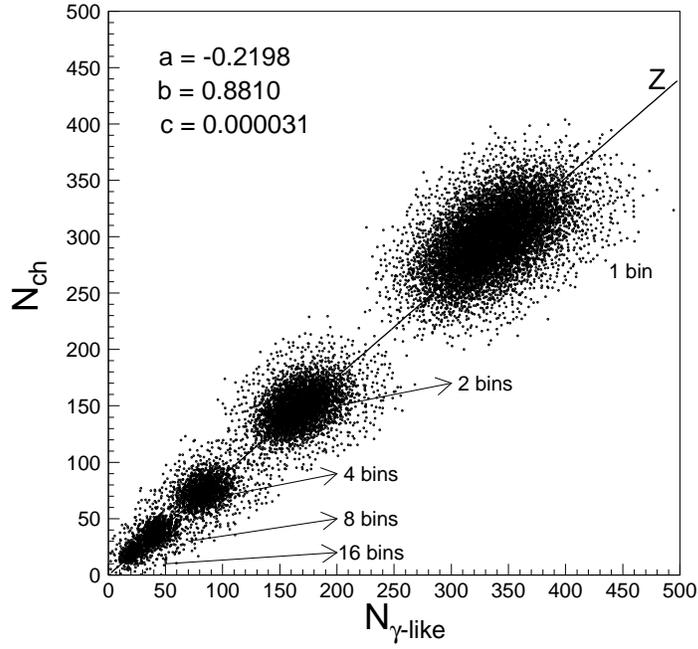}
\caption{ The event-by-event correlation between $N_{\mathrm {ch}}$ and
$N_{\gamma-{\mathrm {like}}}$ for centrality bin=1 
Overlaid on the plot is the common correlation axis (Z-axis) obtained 
for the full distribution by fitting the  $N_{\gamma-{\mathrm {like}}}$ 
and $N_{\mathrm {ch}}$ correlation with a second order polynomial. The
values in the upper left are the coefficients of the polynomial of the form
$N_{\mathrm {ch}}$ = a + b$N_{\gamma-{\mathrm {like}}}$+c$N_{\gamma-{\mathrm {like}}}^{2}$.
For the other three centrality classes, the plots are qualitatively similar.
}
\label{ngam_nch_cor}
\end{center}
\end{figure*}

   The event-by-event correlation between $N_{\gamma-{\mathrm {like}}}$ and
   $N_{\mathrm {ch}}$ has been studied in various $\phi$-intervals by dividing
   the entire  $\phi$-space into 2, 4, 8, and 16 bins. The method of 
   analysis is similar to that described in Refs.~\cite{WA98-3,WA98-12}.
   Fig.~\ref{ngam_nch_cor} shows scatter plot of the correlation between
   $N_{\gamma-{\mathrm {like}}}$ and $N_{\mathrm {ch}}$ 
   for the top centrality bin. The correlation plots for each  
   $\phi$ interval size, including the case of the full interval with 
   no segmentation, are shown. The distributions for the other three
   centrality bins are qualitatively similar. A common correlation axis ($Z$) 
   has been obtained for the full distribution by fitting the  
   $N_{\gamma-{\mathrm {like}}}$ and $N_{\mathrm {ch}}$ correlation with a 
   second order polynomial. The correlation axis with fit parameters is 
   shown in the figure. The distance of separation ($D_{Z}$) between a
   data point and the correlation axis has been calculated with
   the convention that $D_{Z}$ is positive for points below the $Z$-axis.
   The distribution of $D_Z$ represents the relative fluctuations of
   $N_{\gamma-{\mathrm {like}}}$ and $N_{\mathrm {ch}}$ from the
   correlation axis for any chosen $\phi$ bin size. In order to compare the 
   fluctuations for different $\phi$ bins on a similar footing, a scaled
   variable, $S_{Z} = D_Z/s(D_Z)$, is used where $s(D_Z)$ represents
   the rms deviation of the $D_{Z}$ distribution for VENUS events 
   analyzed in the same manner. 
   The $D_Z$ distributions of 
   data, mixed events and the simulated events for a given centrality and 
   $\phi$ bin size are all scaled by the same $s(D_Z)$ corresponding to 
   the VENUS events for the  respective centrality and azimuthal bin size.
   The presence of events with localized
   fluctuations in $N_{\gamma-{\mathrm {like}}}$ and $N_{\mathrm {ch}}$, 
   at a given $\phi$ bin, is expected to result in a broader
   distribution of $S_Z$ compared to those for normal events. 
   Comparing the rms deviations of the  $S_Z$  
   distributions of data, mixed events (to be discussed later), 
   and VENUS events 
   may allow to infer the presence of non-statistical localized 
   fluctuations.

\subsection{Multi-resolution DWT analysis}

   A multi-resolution analysis using discrete wavelet
   transformations (DWT) \cite{amara} has been shown to be quite
   powerful in the search for localized domains of
   DCC \cite{huang,dccflow,dccstr}. The beauty of the DWT technique 
   lies in its power to analyze a spectrum at different resolutions 
   with the ability to identify fluctuations present at any scale.
   This method has been utilized very successfully in many fields including 
   image processing, data compression, turbulence, human vision, radar,
   and earthquake prediction \cite{amara}. 
   It should be noted that the DWT analysis provides different information 
   than the moment analysis of the previous section. 
   It analyzes the event-by-event distribution in phi space to characterize
   the bin-to-bin fluctuations relative to the average behaviour.

   For the present DWT analysis the full azimuthal region
   is divided into smaller bins in $\phi$, the
   number of bins at a given scale $j$ being $2^j$.
   The input to the analysis is a spectrum of the
   sample function at the smallest bin in
   $\phi$ corresponding to the highest resolution scale, $j_{max}$.
   In the present case the sample function is chosen to be the photon
   fraction, given by:

\begin{equation}
f^\prime(\phi) = {N_{\gamma-{\mathrm {like}}}(\phi)}/
{(N_{\gamma-{\mathrm {like}}}(\phi)+N_{\mathrm {ch}}(\phi))}
\end{equation}

   A multi-resolution analysis has been carried out using the $D-$4 wavelet basis 
   on the above sample function starting with $j_{max}=5$.
   It may be mentioned that there are several families of wavelet bases 
   distinguished by the number of coefficients and the level of iteration; 
   we have used the frequently employed $D-$4 wavelet basis \cite{numeric}.
   The output of the DWT consists of a set of wavelet or father function
   coefficients (FFC) at each scale, from $j=1$,...,($j_{max}-1$).
   The coefficients obtained at a given scale, $j$, are derived from 
   the distribution of the sample function at one higher scale, $j+1$.
   The FFCs quantify  the bin-to-bin fluctuations in
   the sample function at that higher scale relative to the average
   behaviour. The presence of localized non-statistical fluctuations
   will increase the rms deviation of the distribution of FFCs and may 
   result in non-Gaussian tails \cite{huang,dccstr}. 
   The DWT technique as used in this analysis has been demonstrated in our 
   earlier publication~\cite{WA98-12}.
   Once again, comparing the rms deviations of the FFC distributions of data, 
   mixed events, and VENUS events may allow to infer the presence of
   localized fluctuations. The utility of the mixed events and the response
   of the analysis technique is demonstrated in Section~V.

\section{CONSTRUCTION OF MIXED EVENTS}

   It is possible to search for non-statistical fluctuations in 
   the experimental data in a model independent way by comparison 
   of the data with mixed events generated from the data itself.
   Furthermore, it is necessary to isolate the various contributions 
   to the fluctuations and to understand all detector related effects 
   in the data. This has been done by generating different types
   of mixed events which preserve the global multiplicity correlation
   between $N_{\gamma-{\mathrm {like}}}$ and $N_{\mathrm {ch}}$. 
   Fluctuations in the ratio of $N_{\gamma}$
   to $N_{\mathrm {ch}}$ can arise due to fluctuations in $N_{\gamma}$ only,
   fluctuations in $N_{\mathrm {ch}}$ only, or fluctuation in both
   $N_{\gamma}$ and $N_{\mathrm {ch}}$. Furthermore, the fluctuations in 
   $N_{\gamma}$ and $N_{\mathrm {ch}}$  may be correlated event-by-event, as
   nominally expected in the case of DCC formation. 
   Each of these possibilities is investigated through the construction
   of four different kinds of mixed events. 
   The method of construction of these mixed events and the type
   of fluctuations they probe are described next.

\subsection{Maximally Mixed Events}

   The first set of mixed events, referred to as M1 events,
   are constructed to remove all correlations to the greatest 
   extent possible to provide a baseline for comparison to the real 
   event data.
   They were generated by mixing hits in both the photon and charged
   particle detectors separately but still satisfying the global 
   $N_{\gamma-{\mathrm {like}}}$-$N_{\mathrm {ch}}$ correlation of the real event in the 
   full acceptance. 
   This means that on an event-by-event basis the total photon multiplicity 
   and charged particle multiplicity of the mixed event were identical to 
   those of the real event to which it corresponds. 
   Thus, the scatter plot of the mixed events is
   identical to the real events for the single bin case shown in  
   Fig.~\ref{ngam_nch_cor}.
   The idea is to constrain the mixed events to be identical to real events globally
   and then compare them to the real data in localized regions of phase
   space to search for indications of non-statistical localized fluctuations 
   in the data.
   
   The M1 type of mixed events were constructed from the pool of all
   photon-like and charged particle hits, in which the hit position ($\eta$
   and $\phi$) and  event information (event number and total multiplicity) 
   was kept for both the photon-like and charged particle hits.
   For a given real event measured to have multiplicities 
   $N_{\gamma-{\mathrm {like}}}$ and $N_{\mathrm {ch}}$ in the full acceptance region, a
   mixed event was constructed by randomly selecting $N_{\gamma-{\mathrm {like}}}$ 
   photon-like hits 
   from the pool of photon-like hits and  $N_{\mathrm {ch}}$ charged particle
   hits from the pool of charged particle hits. 
   This procedure was repeated for each real
   event.  Care was taken such that no two hits from the same real
   event were used in the construction of a mixed event. 
   Also, for the mixed events, hits within either the SPMD or PMD detector 
   were not allowed to lie within the two-track resolution of that detector.
   In brief, the M1 mixed events randomly
   distribute the hits in each individual detector but keep the global
   correlation between the $N_{\gamma-{\mathrm {like}}}$ and $N_{\mathrm {ch}}$ 
   multiplicity.
   They provide a maximally randomized sample of PMD and SPMD hits. 
   Comparisons of such mixed events to real
   events will be most sensitive to the presence of localized
   fluctuations. However, in themselves
   they would not isolate the source of fluctuations as being
   due to $N_{\gamma}$ and/or $N_{\mathrm {ch}}$, or 
   correlations between $N_{\gamma}$ and $N_{\mathrm {ch}}$.

\subsection{Minimally Mixed Events}

   A second type of mixed events, referred to as M2 mixed events, were
   constructed to investigate the presence of
   correlated event-by-event fluctuations between $N_{\gamma}$ and
   $N_{\mathrm {ch}}$. These mixed events had a minimal amount of randomization
   since they were generated by mixing the photon hits taken unaltered from one
   event with the
   charged particle hits taken unmodified from another event. 
   As with the M1 mixed events, the global  
   $N_{\gamma-{\mathrm {like}}}$-$N_{\mathrm {ch}}$ multiplicity 
   correlation was maintained to be
   exactly the same as the data. 
   To construct such mixed events, for a given real event measured 
   to have multiplicities $N_{\gamma-{\mathrm {like}}}$ and $N_{\mathrm {ch}}$ in the full
   acceptance region, a mixed event was constructed by keeping the 
   PMD $N_{\gamma-{\mathrm {like}}}$ portion of the event intact, but combining it 
   with the unaltered SPMD portion of a different randomly selected 
   event, but constrained to have almost the same charged particle 
   multiplicity $N_{\mathrm {ch}}$. This procedure was repeated for each real event.
   In brief, this type of mixed event keeps the  
   event-by-event hit structure in each detector  
   identical to that in real events. Thus, such mixed events keep the 
   individual localized fluctuations present in $N_{\gamma}$ or $N_{\mathrm {ch}}$, 
   but remove the event-by-event localized correlated fluctuations between them.
   Comparison of such mixed events to real
   events may reveal the presence of correlated localized
   fluctuations between $N_{\gamma}$ and $N_{\mathrm {ch}}$.

\subsection{Partially Mixed Events}

   Intermediate between the M1 and M2 types of mixed events are a third and 
   fourth type of partially mixed event, referred to as M3-$\gamma$ and 
   M3-${\mathrm {ch}}$ mixed events. These were constructed to provide 
   information regarding the contribution to 
   the localized fluctuations in the $N_{\gamma}$ to $N_{\mathrm {ch}}$
   ratio from the individual observables ($N_{\gamma}$ and
   $N_{\mathrm {ch}}$). They were generated from real events by
   mixing hits in one of the detectors (following the procedure for
   construction of M1 mixed events) and keeping the hit structure of
   the event in the other detector intact. M3-$\gamma$ mixed events
   correspond to the 
   case where the hits within the photon detector are unaltered while 
   the hits in the charged particle detector are mixed. Similarly in 
   M3-$\mathrm {ch}$ mixed events the hits in the charged particle 
   detector were unaltered and the hits in the photon detector were mixed.
   In each type of mixed event the global $N_{\gamma-{\mathrm {like}}}$--$N_{\mathrm {ch}}$ 
   correlation is maintained as in the real event. The two track resolution 
   in the detectors where the hits are mixed is kept identical to that in 
   real events. The total number of mixed events is the same as the number of  
   real events. Comparison of such mixed events to real
   events and the other types of mixed events will reveal 
   the presence of localized fluctuations in $N_{\gamma}$ or 
   $N_{\mathrm {ch}}$ separately.

   A summary of the different sources of fluctuations 
   in the ratio of $N_{\gamma}$
   to $N_{\mathrm {ch}}$ probed by each of the types
   of mixed events is given in Table~1.
   
\begin{table}
\caption{\label{tab:table1} Type of fluctuations preserved by
various mixed events}
\begin{ruledtabular}
\begin{tabular}{cccccc}
Fluctuation& &  &Mixed & Event & \\ 
 & &M1&M2&M3-$\mathrm {ch}$& M3-$\gamma$ \\
\hline
$N_{\gamma}$-only& &No & Yes & No & Yes \\
$N_{\mathrm {ch}}-only$& &No & Yes & Yes & No \\
correlated $N_{\gamma}$-$N_{\mathrm {ch}} $& &No & No & No &No \\ 

\end{tabular}
\end{ruledtabular}
\end{table} 

\section{DEMONSTRATION OF ANALYSIS METHOD}

In this section we wish to demonstrate the analysis method and
illustrate how the relationship of the measured result to that
obtained with the various mixed events can be used to provide an
essentially model-independent signature of DCC formation.
To demonstrate the analysis method and the potential to observe DCC
event formation, we have applied
the DWT analysis to a simple DCC-like model. The analysis is applied to
``real'' DCC events from the model as well as the various types of mixed events
described in the previous section constructed from the model DCC
events. Since event generators which include DCC formation do not
exist, we have
implemented a simple DCC model in which 
localized non-statistical $N_{\gamma}$-$N_{ch}$ fluctuations have been 
introduced by modification of the output of the VENUS event 
generator. To implement the fluctuations,  
the final state pions within a localized $\eta$--$\phi$ region from
VENUS are interchanged pairwise ($\pi^{+}\pi^{-} \leftrightarrow
\pi^{0}\pi^{0}$) according to the DCC probability
distribution, $P(f) = 1 / 2 \sqrt{f}$. 
The fluctuations were generated over a localized region of
$\eta=3-4$ and a $\Delta\phi$ interval of $90^{0}$.   The 
$\pi^{0}$'s were then allowed to decay. The resulting events were
then passed through the WA98 detector response simulation.
The DCC events in the simple model used here give rise to 
an anti-correlation between $N_{\gamma}$ and $N_{\mathrm ch}$. 
It also results in non-random fluctuations in both $N_{\gamma}$ 
and $N_{\mathrm ch}$ individually.   
Since the
probability to produce events with localized charged-neutral
fluctuations is unknown, ensembles of events, here referred to as 
``nDCC events'', were produced as a mixture of normal VENUS events 
and events with localized fluctuations. The fraction of events
with localized DCC-like fluctuations in each nDCC sample was varied
as a parameter to be studied. 

By using VENUS
events as the basis to introduce the DCC effect, it is insured that the
general features of the event, including the
multiplicity, composition, and momenta of the produced particles, as
well as correlations in the particle multiplicities due to impact
parameter variation, are reasonably well described. 
Also the GEANT simulation of the detector response
to these events insures that other effects such as photon conversions
and the response of the PMD to hadrons, which might affect the observed
multiplicities and the observed correlations, are also taken into
account. While the assumption of a 100\% DCC contribution over an 
interval of fixed size in $\eta-\phi$ is certainly a gross simplification, it
provides a well-defined reference to gauge the potential for DCC
observation. Other assumptions, such as a varying fraction, which might
also be momentum dependent (since DCCs are expected to be a low p$_T$
phenomena), and different or varying sizes might be more realistic. However,
without clear theoretical guidance we have chosen this very simple
model as a reasonable and  well-defined reference.

\begin{figure}
\begin{center}
\includegraphics[scale=0.5]{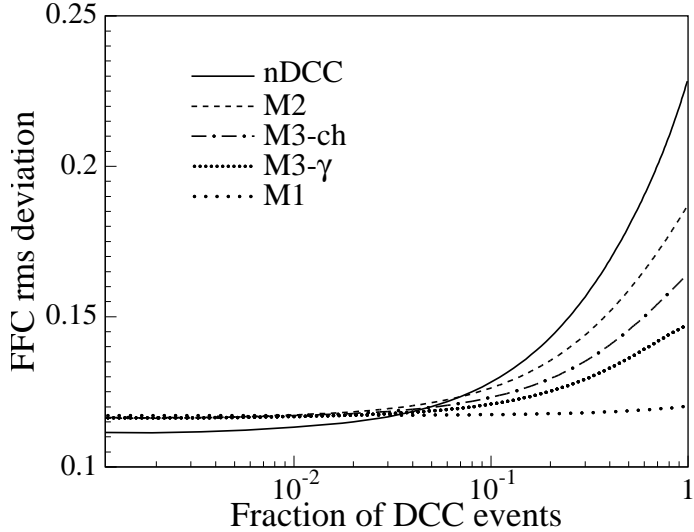}
\caption {\label{rms}
The rms deviations of the FFC distributions for
simulated VENUS events containing a variable fraction of 
localized DCC-like events with DCC extent $\Delta\phi_{DCC}=90^\circ$,
as a function of that fraction. Results are also shown  
for various mixed events constructed from those events.
}
\end{center}
\end{figure}

The DWT analysis was carried out on an ensemble of nDCC events and
their corresponding mixed event sets created from each set of nDCC events. 
By ensemble of nDCC events we mean 
sets of events having different percentages of events with localized
fluctuations. The percentage varied from zero, that is, normal 
VENUS events with no localized fluctuations, 
to an event set where all events had localized charged-neutral
fluctuations.  For every nDCC
event set the four sets of mixed events (M1,M2, M3-$\gamma$, and
M3-ch) were constructed from the nDCC events and analyzed. 
The rms deviations of the FFC distributions 
for each set of nDCC events and corresponding mixed event sets were
obtained. The results for scale $j=1$ 
are plotted in Fig.~\ref{rms} as a function of
the percentage of localized DCC-type events in the nDCC event set. 

It is seen that the rms deviations of the FFC distributions of the 
nDCC events increase as 
the percentage of events with localized fluctuations in
$N_{\gamma}$-$N_{\mathrm ch}$ increases. This is the expected behaviour 
and demonstrates the linear response of the DWT to 
the DCC events when the frequency of events with fluctuations
increases. 
On the other hand, the rms deviations of the FFC
distributions of the M1 type mixed events created from the nDCC events
are found to be independent of the percentage of events having localized
fluctuations.
This is also the expected behaviour and demonstrates that the M1 mixed
events can be used as a baseline from which to
deduce the presence of fluctuations in a model-independent manner. 
However, the deviation of the
real events from the M1 mixed events does not inform about the relative
contributions of the individual $N_{\gamma}$ and $N_{\mathrm ch}$
fluctuations. The rms deviations of the FFC distributions of the M3 
mixed events are found to be intermediate to those obtained for nDCC
events and M1 mixed events. They indicate the separate contributions 
of the $N_{\gamma}$ or $N_{\mathrm ch}$ fluctuations alone to the
ratio.  The rms
deviations of the M2 mixed events are higher than those of M1 and
M3 mixed events. That is because the M2 mixed events 
keep the separate contributions of both the $N_{\gamma}$ and $N_{\mathrm ch}$  
fluctuations. However, the rms deviations of the
M2 mixed events are consistently below those for the nDCC events. This is
because the M2 events randomize the  
correlations between $N_{\gamma}$ and $N_{\mathrm ch}$. The difference
between the M2 mixed events and the nDCC events indicates the 
presence of the DCC-like correlated $N_{\gamma}$-$N_{\mathrm ch}$
fluctuations. The relative pattern of rms values for the real events
and the various mixed events constructed from those real events 
seen in Fig.~\ref{rms} provides
a rather unambiguous model-independent signature for DCC-like fluctuations.
Similar relative pattern of the rms deviations of the $S_{Z}$ distributions
for mixed events and simulated events were also observed for $N_{\gamma}$-$N_{\mathrm ch}$ correlation analysis.
In particular, the observation of fluctuations in real events that are greater than
the M2 mixed events would constitute what might be called ``smoking gun'' 
evidence for DCC formation.
Conversely, the lack of a difference between real events and  
M2 mixed events would indicate
the lack of DCC-like correlated charged-neutral fluctuations.

Figure~\ref{rms} also demonstrates an effect which must be taken into
account when comparing the measured result to the mixed events. 
For nDCC events with vanishing fraction of events with fluctuations,
which is to say for normal VENUS events, 
it is seen that the rms deviations of the FFC
distributions for all types of mixed events are higher than those for
the nDCC events. In the VENUS simulations this is due to the presence of 
correlations between $N_{\mathrm ch}$ and $N_{\gamma-{\mathrm like}}$.
These are primarily due to residual impact parameter correlations
(see Fig.~\ref{ngam_nch_cor}) as well as
due to the charged particle contamination in the
$N_{\gamma-{\mathrm like}}$ data sample whereby the charged particles
register in both the PMD and SPMD (see Section~II.C)~\cite{bedanga}. These
correlations are removed by the event mixing procedure which results in a
larger rms deviations for the mixed events.  The real data is presumed
to have similar residual correlations as observed in the VENUS
simulations. In order to correct for the effects of these non-DCC
correlations, all mixed event rms
values constructed from real events have been rescaled by the 
percentage difference between the rms
deviations of the VENUS distributions and those of the corresponding
VENUS mixed events, as also discussed in Ref.~\cite{WA98-12}. 

For nDCC
events with larger percentages of DCC-like events, the anti-correlation between
$N_{\mathrm ch}$ and $N_{\gamma-{\mathrm like}}$ overcomes the
correlations between $N_{\mathrm ch}$ and $N_{\gamma-{\mathrm like}}$
and hence the rms deviations of the
FFC distributions of nDCC events become greater than those of the mixed
events, despite the other correlation effects.

\section{RESULTS AND DISCUSSION}

   In the analysis of  experimental data, the results 
   from the measured data are compared with simulated and mixed events.
   Below we discuss the results obtained from such a comparison using
   two different analysis methods discussed earlier.

\subsection{$N_\gamma$ versus $N_{\mathrm {ch}}$ correlation results}

\begin{figure*}
\begin{center}
\includegraphics[scale=0.4]{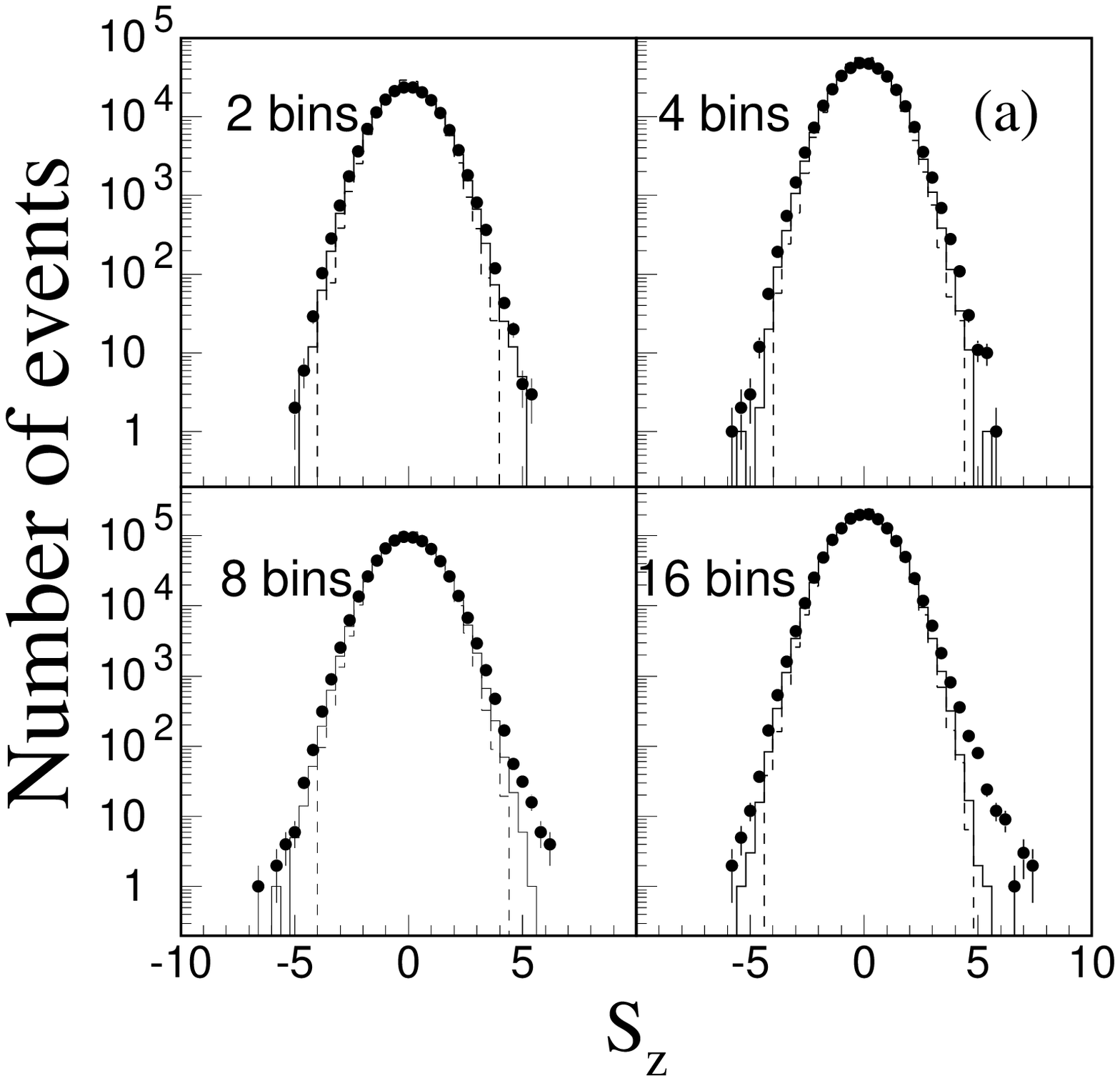}
\includegraphics[scale=0.4]{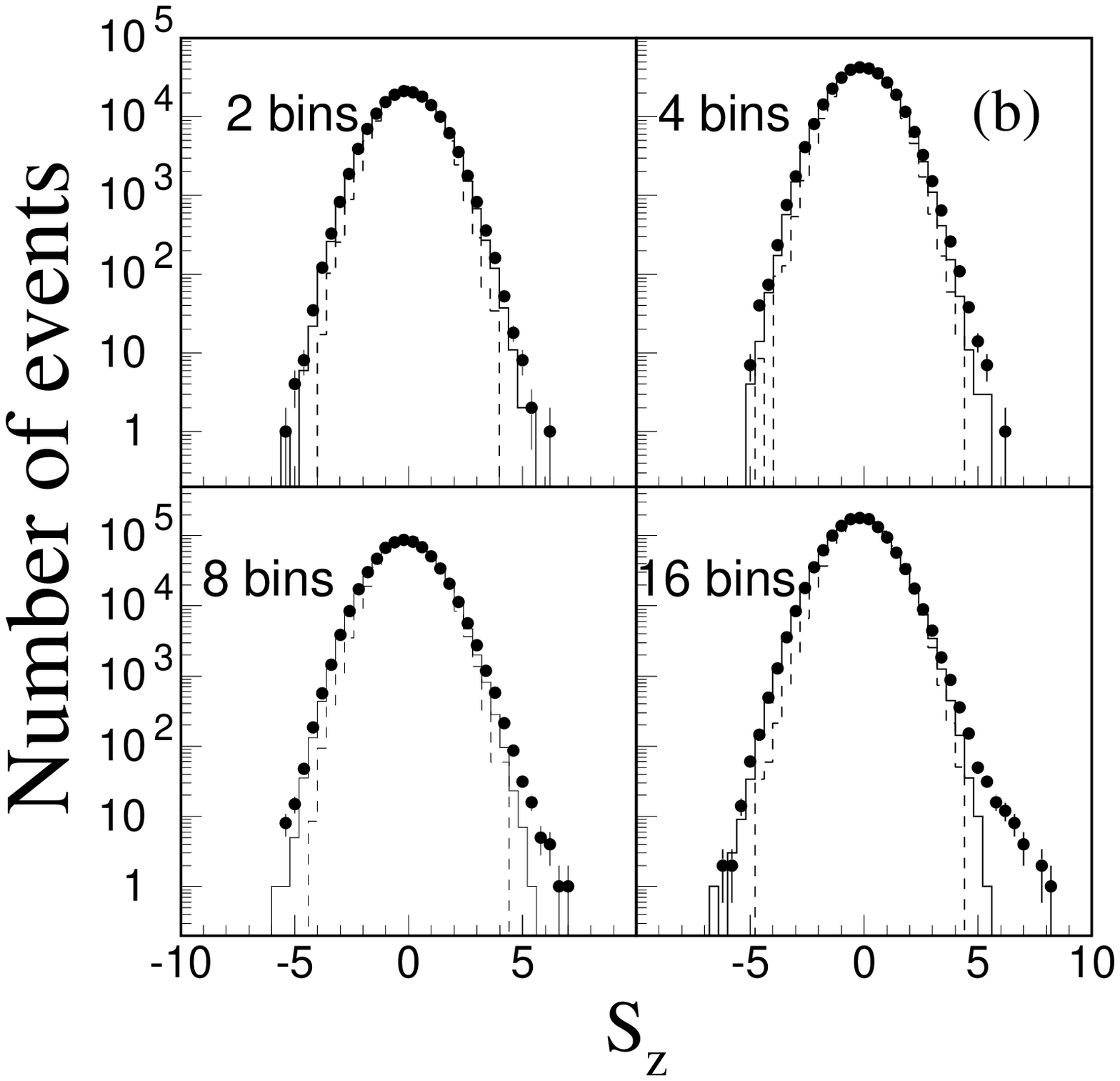}
\includegraphics[scale=0.4]{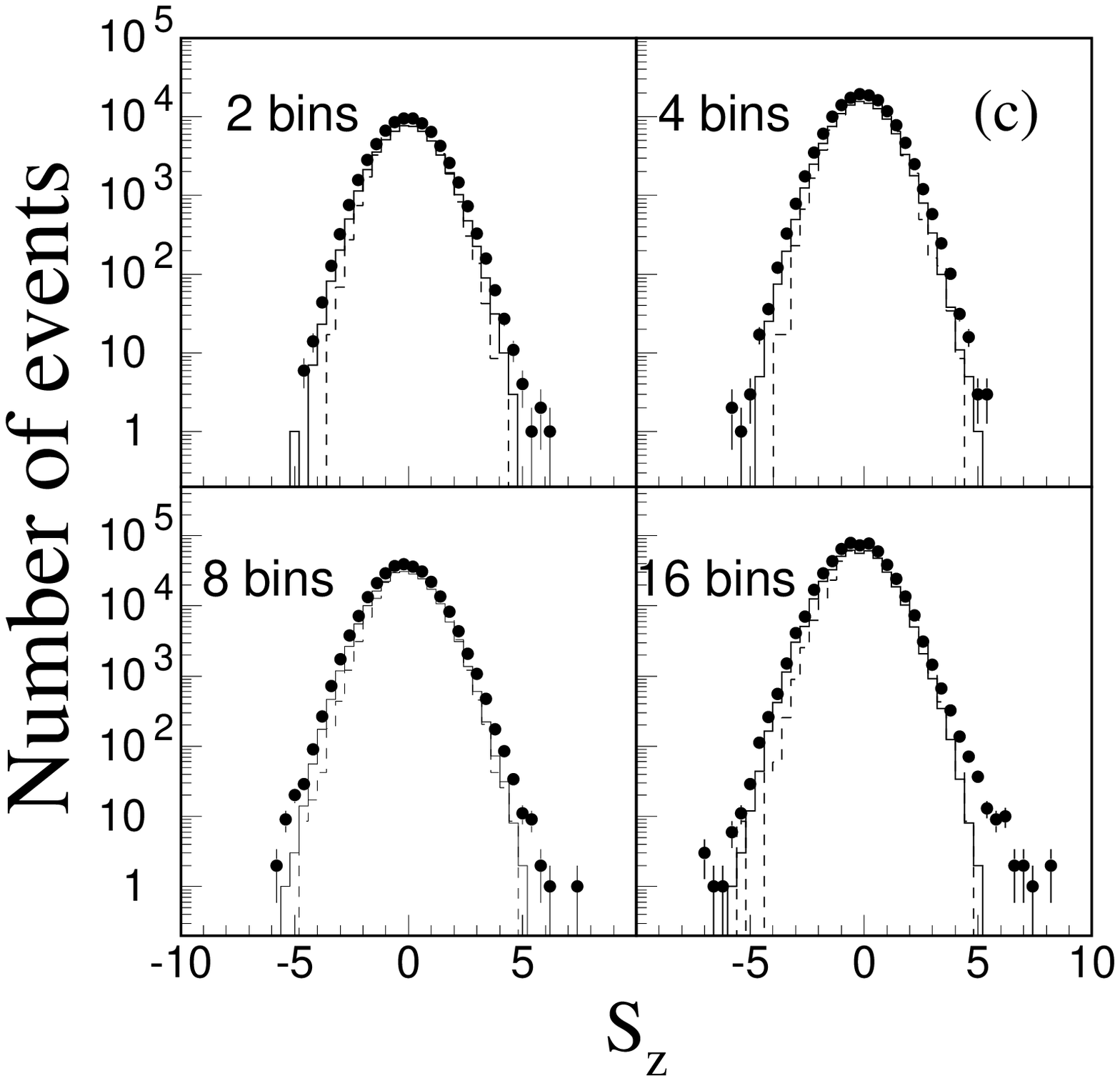}
\includegraphics[scale=0.4]{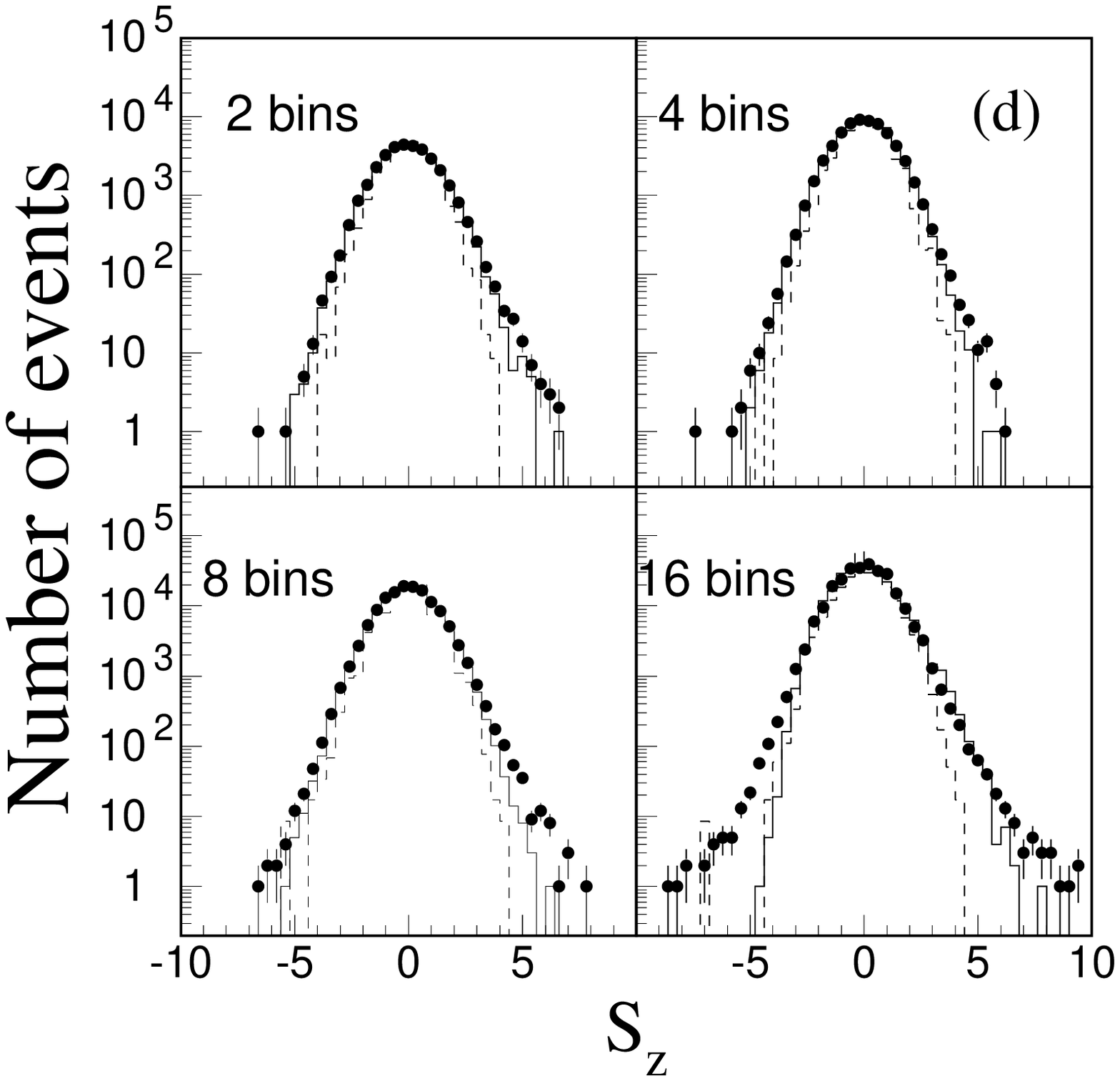}
\caption{The $S_Z$ distributions for data
(solid circles), mixed events (solid histogram), and 
simulated events (dashed histogram)  for the four centrality bins
(a) centrality-1 ($0-5\%$), (b) centrality-2 ($5-10\%$), (c) 
centrality-3 ($15-30\%$), (d) centrality-4 ($45-55\%$).
}
\label{sz}
\end{center}
\end{figure*}

\begin{figure*}
\begin{center}
\includegraphics[scale=0.4]{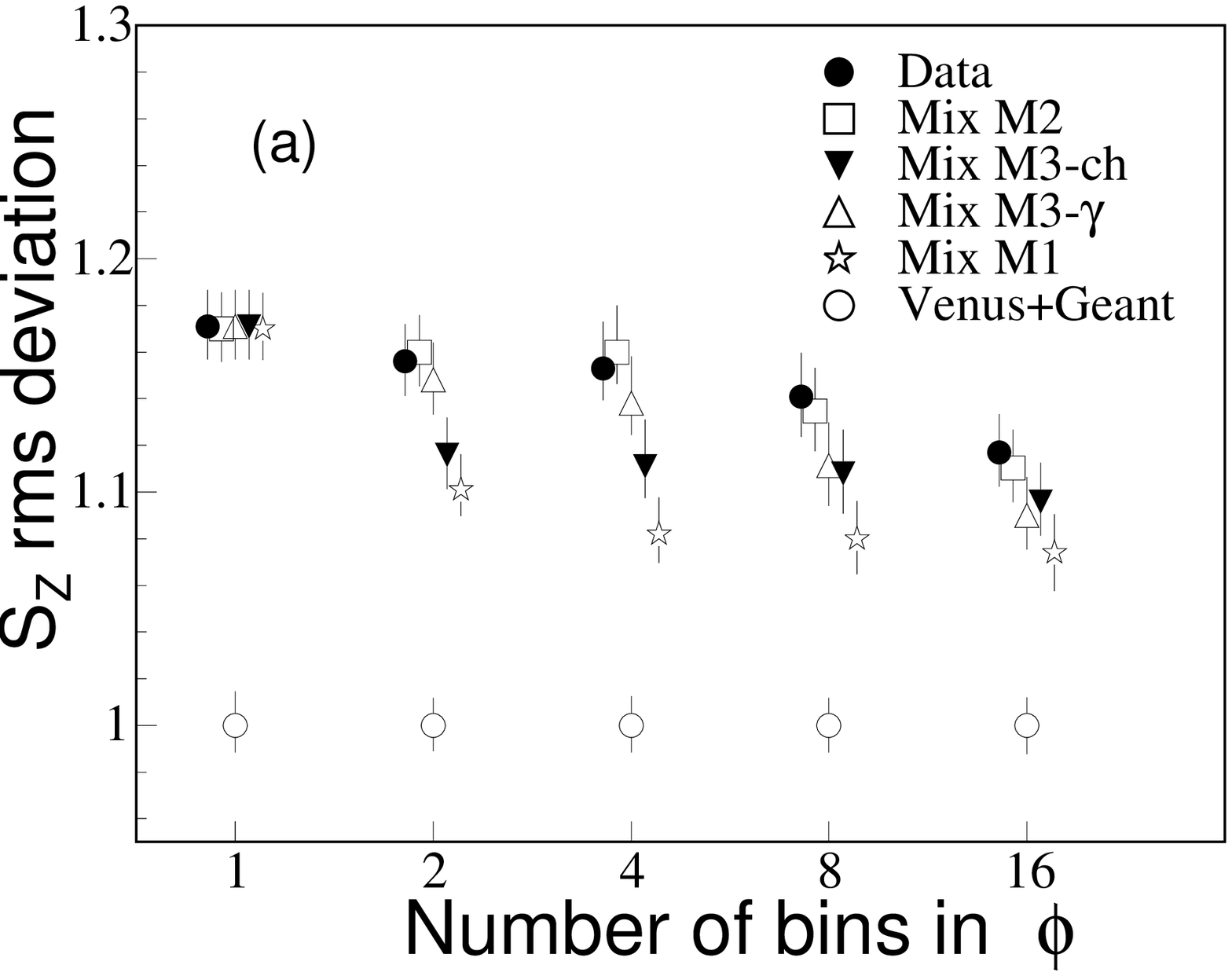}
\includegraphics[scale=0.4]{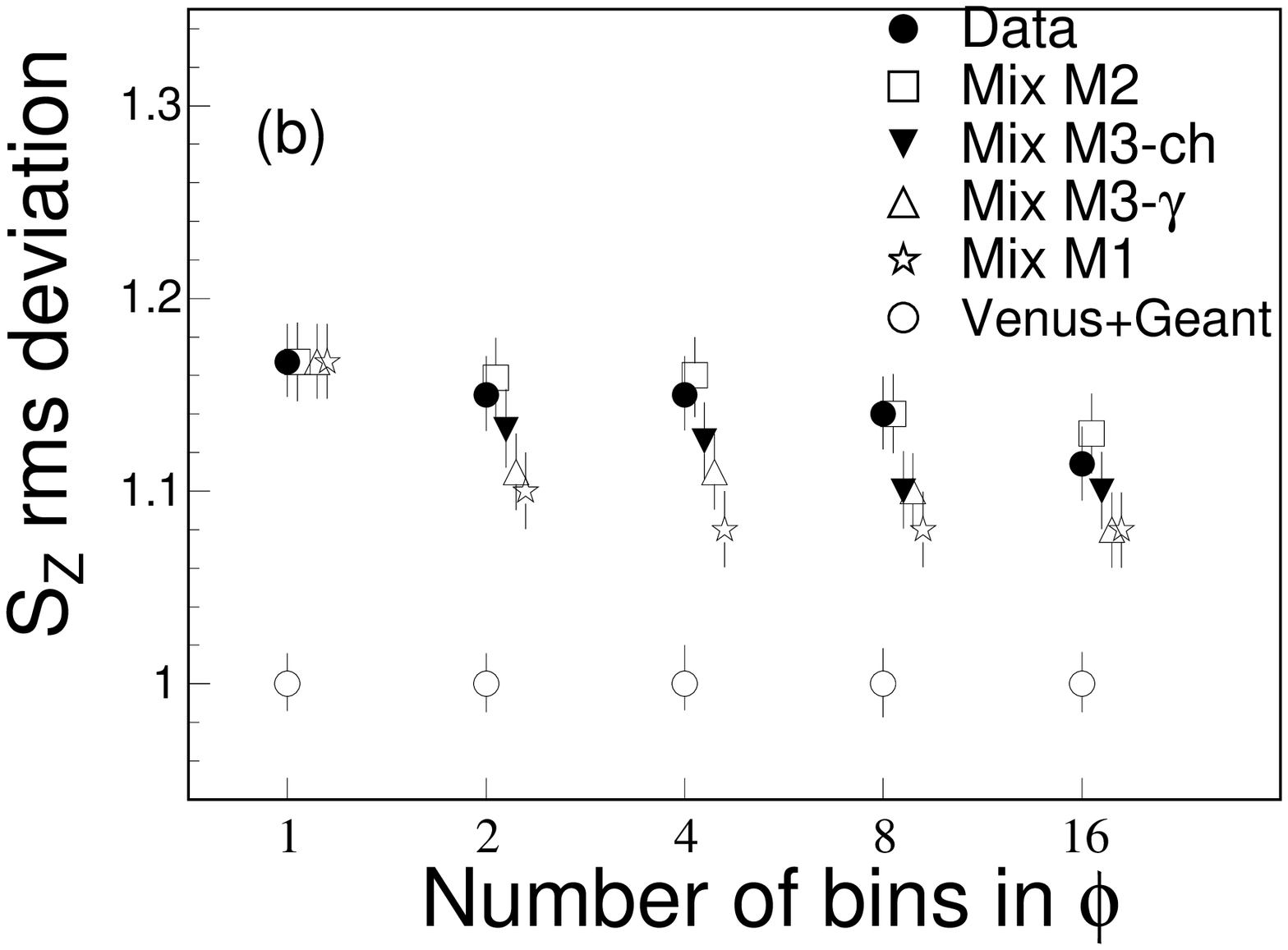}
\includegraphics[scale=0.4]{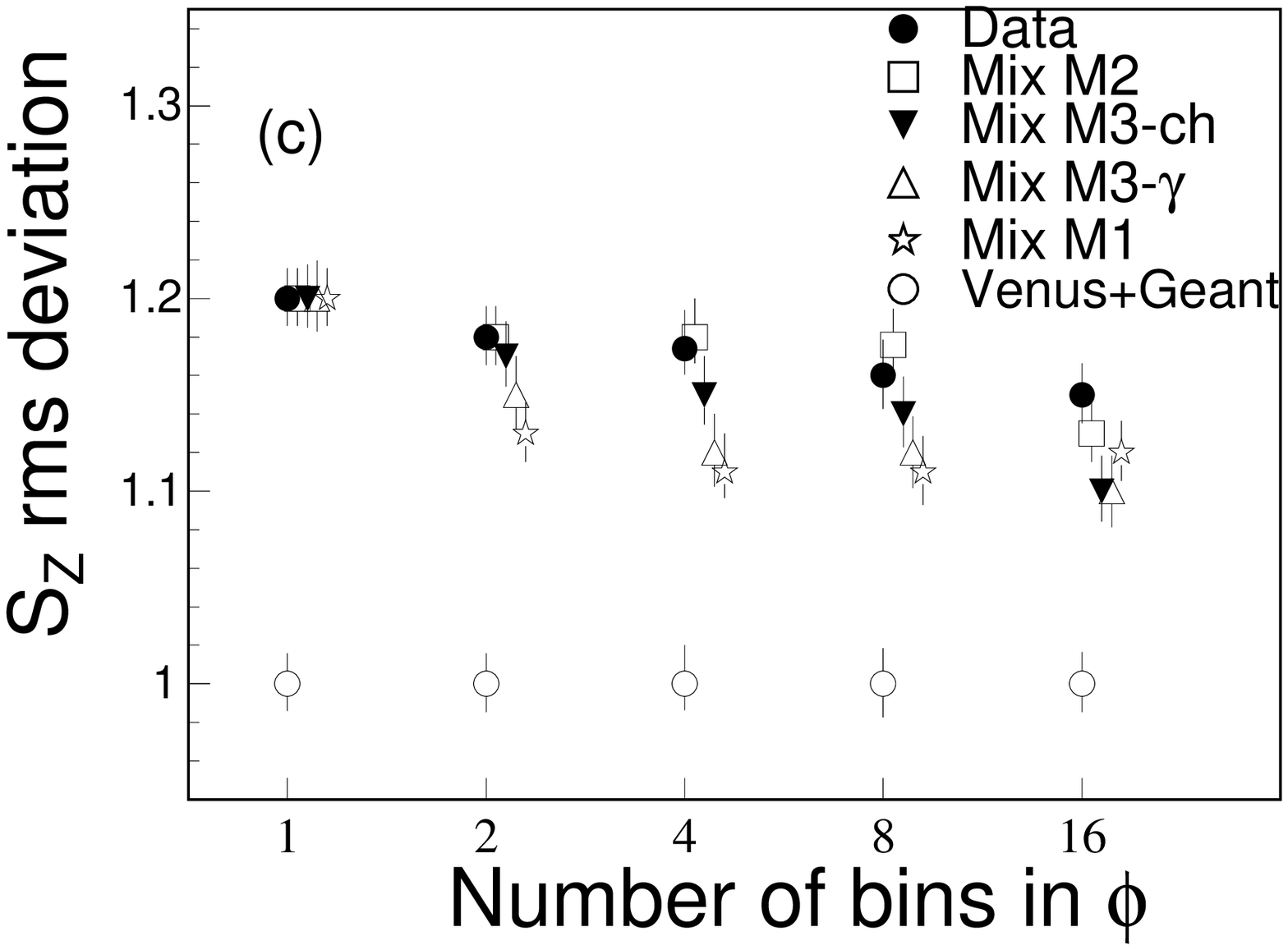}
\includegraphics[scale=0.4]{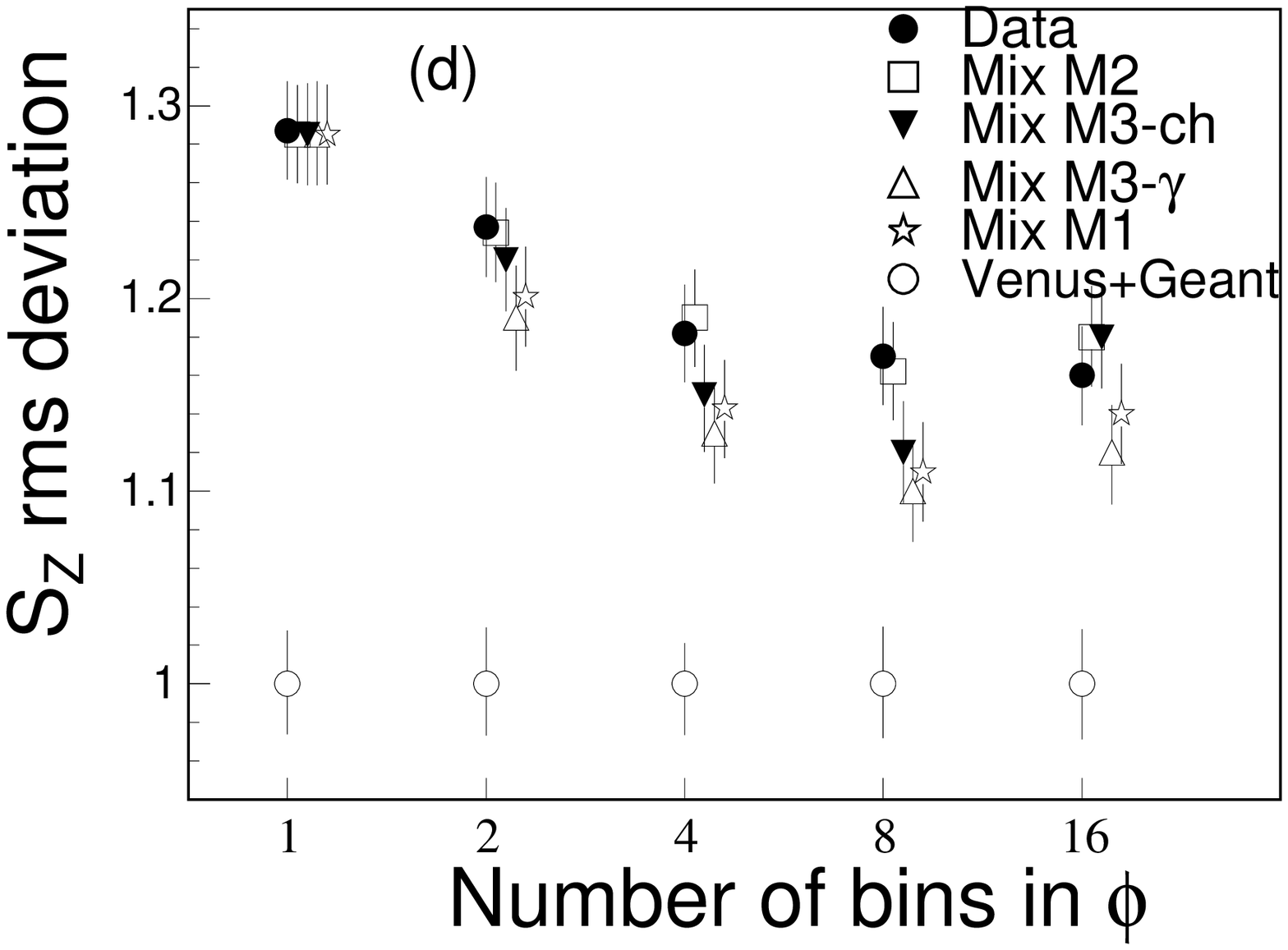}
\caption{The RMS deviations of the $S_Z$ distributions for data,
various mixed events, and simulated events for the four centrality bins
(a) centrality-1 ($0-5\%$), (b) centrality-2 ($5-10\%$), (c) 
centrality-3 ($15-30\%$), (d) centrality-4 ($45-55\%$).
}
\label{sz_rms}
\end{center}
\end{figure*}

   The $S_{Z}$ distributions  calculated for different
   $\phi$ bin sizes are shown in Fig.~\ref{sz} for data, M1,
   and VENUS events, for the four different centrality selections. 
   The distributions for the other types of mixed events 
   are not shown for clarity of presentation. The small differences
   in the $S_Z$ distributions have been quantified in terms of the
   corresponding rms deviations, of these distributions
   shown in Fig.~\ref{sz_rms}. The statistical errors on the values
   are small and are within the size of the symbols.
   The bars represent statistical and systematic errors added in
   quadrature. The various sources of systematic error have  been
   discussed in an earlier section. An additional systematic error  
   from the fit errors associated with the determination of the 
   correlation axis ($Z$) is also included.  In general, the
   width of the $S_Z$ distribution increases from the most central to
   less central event selections and decreases with decreasing bin
   size. 

   The rms deviations of the $S_Z$ distribution 
   for the VENUS simulated events are $1$ (by definition) for all
   centrality and all bins in azimuth by definition of $S_{Z}$ and 
   are significantly
   different from the measured results. This is primarily because the global
   particle multiplicity as predicted by VENUS and those measured in the 
   experiment within the coverage of the detectors are not same.

   As seen in Fig.~\ref{sz_rms}, the widths of the $S_Z$ distributions for 
   mixed events closely follow those of the data.
   The mixed events have been constructed such that
   the global 
   $N_{\gamma-{\mathrm {like}}}$ ${\mathrm {vs.}}$ $N_{\mathrm {ch}}$
   multiplicity correlations are maintained.
   Therefore the rms deviations of the data and the mixed event reference 
   are the same in the first $\phi$ bin of Fig.~\ref{sz_rms} by construction.
   Some correlations
   between $N_{\mathrm {ch}}$ and $N_{\gamma-{\mathrm {like}}}$ are 
   expected, mostly as a
   result of the charged particle contamination in the
   $N_{\gamma-{\mathrm {like}}}$ data sample, but are removed by
   the event mixing procedure and thereby result in a small difference between
   the real and mixed events, as seen in the analysis of the VENUS
   events, discussed earlier. 
   All of the mixed event $S_{Z}$ distribution rms values (
   Fig.~\ref{sz_rms}) have therefore been rescaled by the percentage
   difference between the rms deviations of the VENUS  distributions and
   those of the corresponding VENUS mixed events for each centrality 
   class (Ref.~\cite{WA98-12}).

   The rms deviations of $S_{Z}$ distributions of the M2 mixed events are found
   to agree with those of the experimental data within errors for all
   four centrality classes and for all azimuthal bin sizes. 
   This indicates the absence of event-by-event localized correlated 
   fluctuations in $N_{\gamma-{\mathrm {like}}}$ and $N_{\mathrm {ch}}$,
   such as would be expected for DCC-like
   fluctuations. On the other hand, 
   the rms deviations of the M1 mixed events are found 
   to be systematically lower than those of the data for 2, 4, and 8 bins in $\phi$ 
   for centrality bins 1, 2, and 3. The results for both types of M3 
   mixed events are found to be intermediate between those of the data and 
   the M1 mixed events. The results indicate the presence of localized 
   fluctuations in the data in 
   both the photon and charged particle multiplicities. 
   For the case of the most peripheral centrality selection (centrality-4),
   the rms deviations of the $S_{Z}$ distributions of data and the
   various mixed events are found to be in close agreement to each other 
   within the quoted
   errors.

\subsection{Multi-resolution DWT analysis results}

\begin{figure*}
\begin{center}
\includegraphics[scale=0.4]{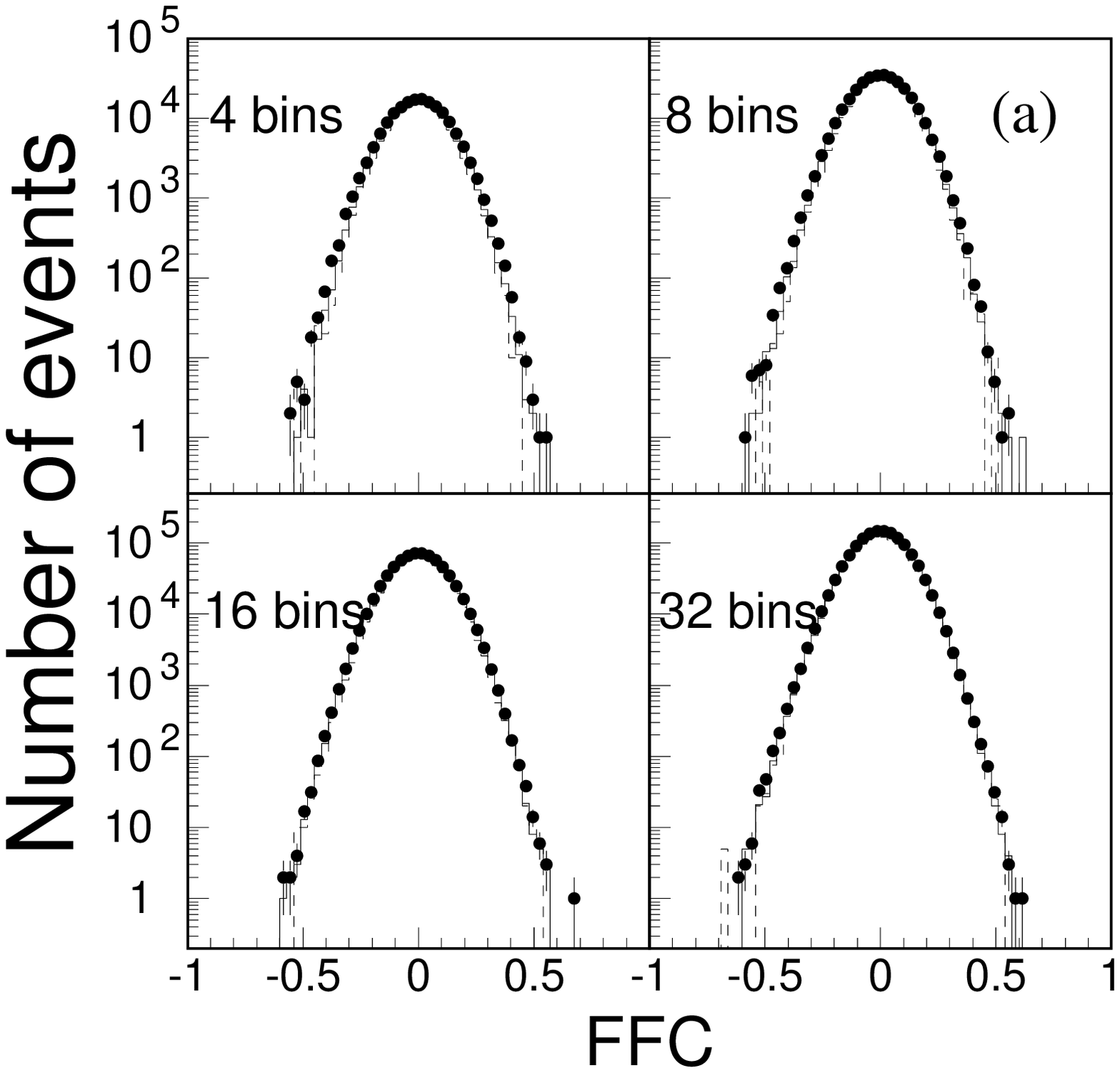}
\includegraphics[scale=0.4]{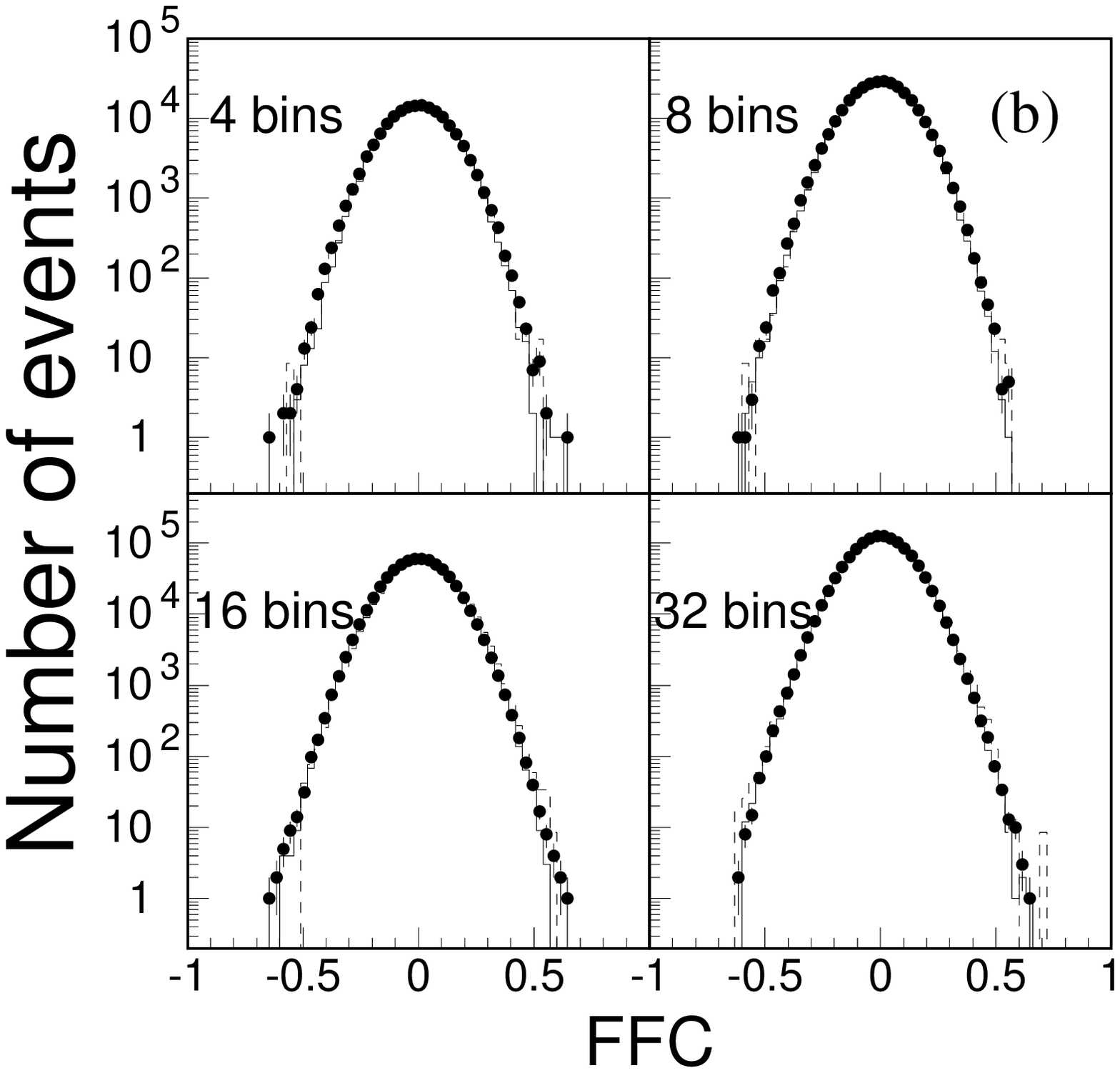}
\includegraphics[scale=0.4]{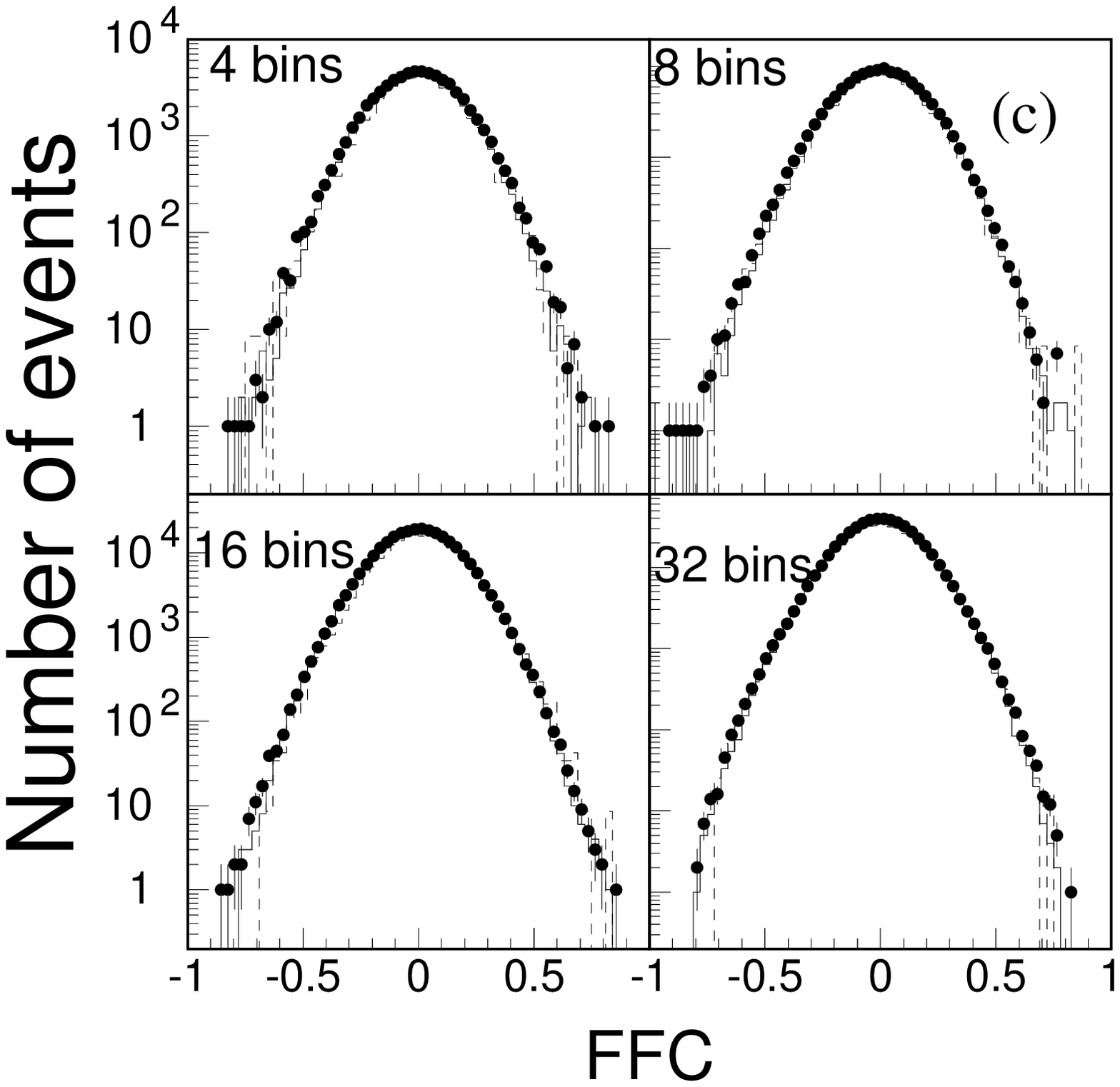}
\includegraphics[scale=0.4]{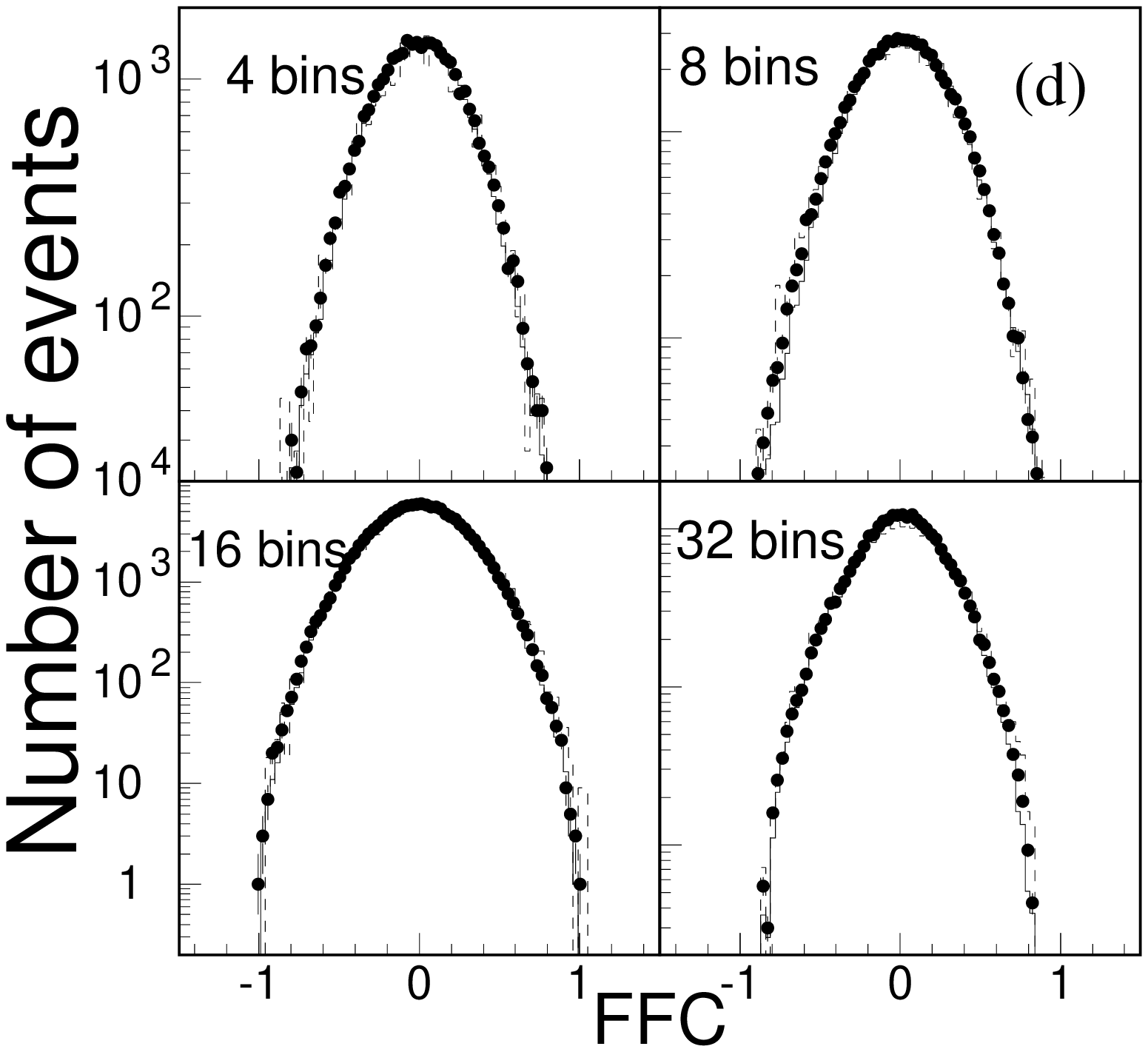}

\caption{The FFC distributions for data
(solid circles), mixed events (solid histogram), and 
simulation (dashed histogram)  for the four centrality bins
(a) centrality-1 ($0-5\%$), (b) centrality-2 ($5-10\%$), (c) 
centrality-3 ($15-30\%$), (d) centrality-4 ($45-55\%$).
}
\label{ffc}
\end{center}
\end{figure*}

\begin{figure*}
\begin{center}
\includegraphics[scale=0.4]{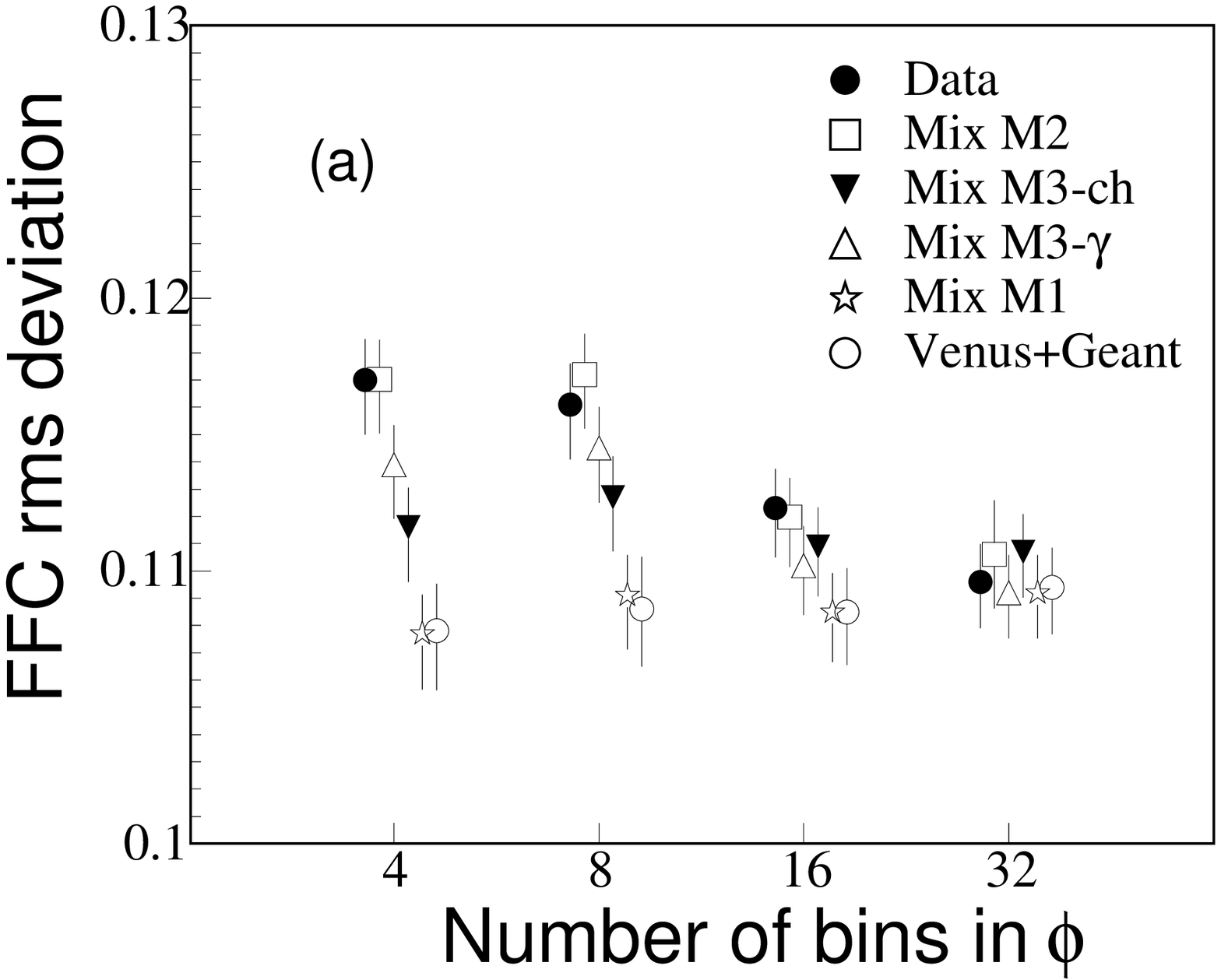}
\includegraphics[scale=0.4]{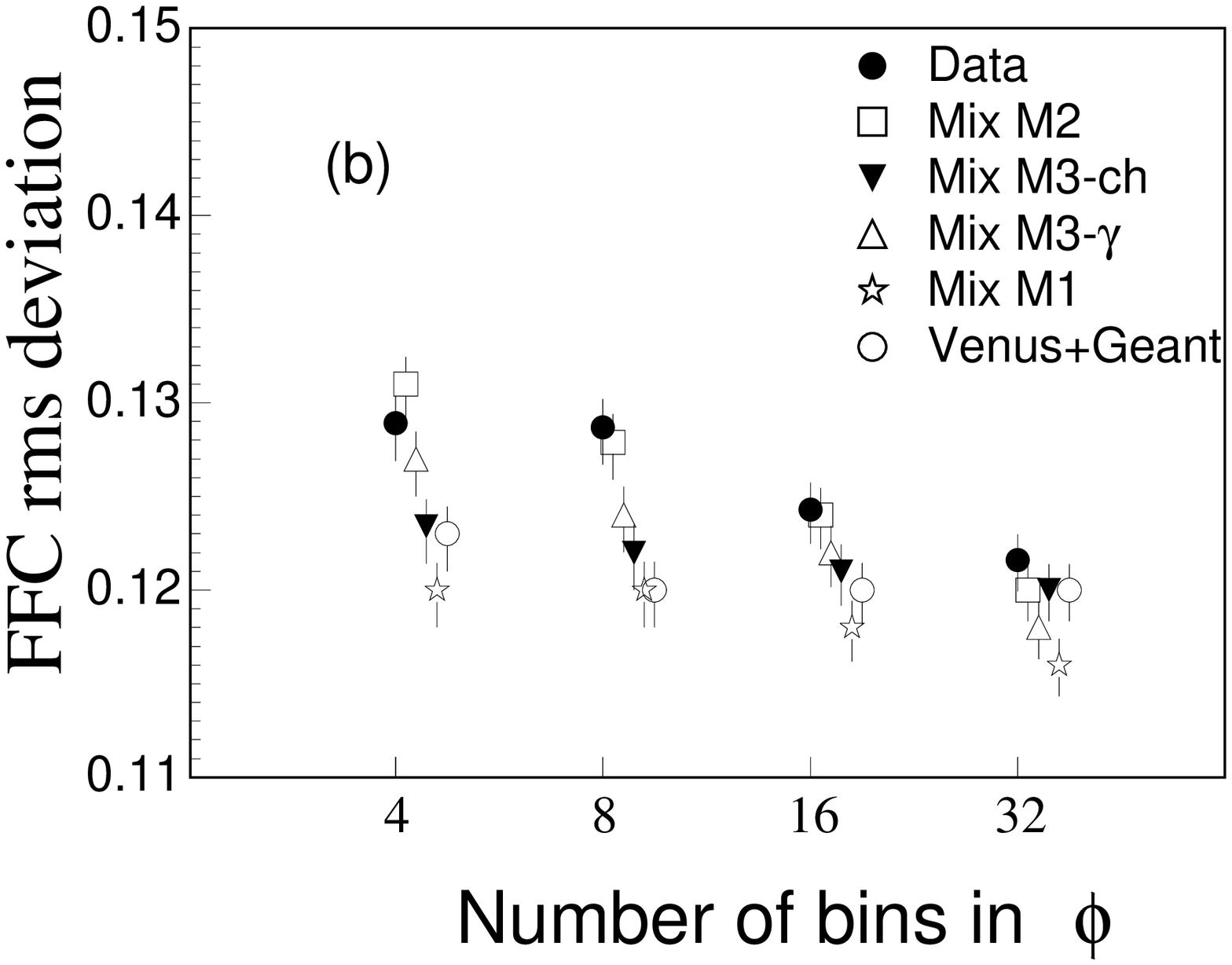}
\includegraphics[scale=0.4]{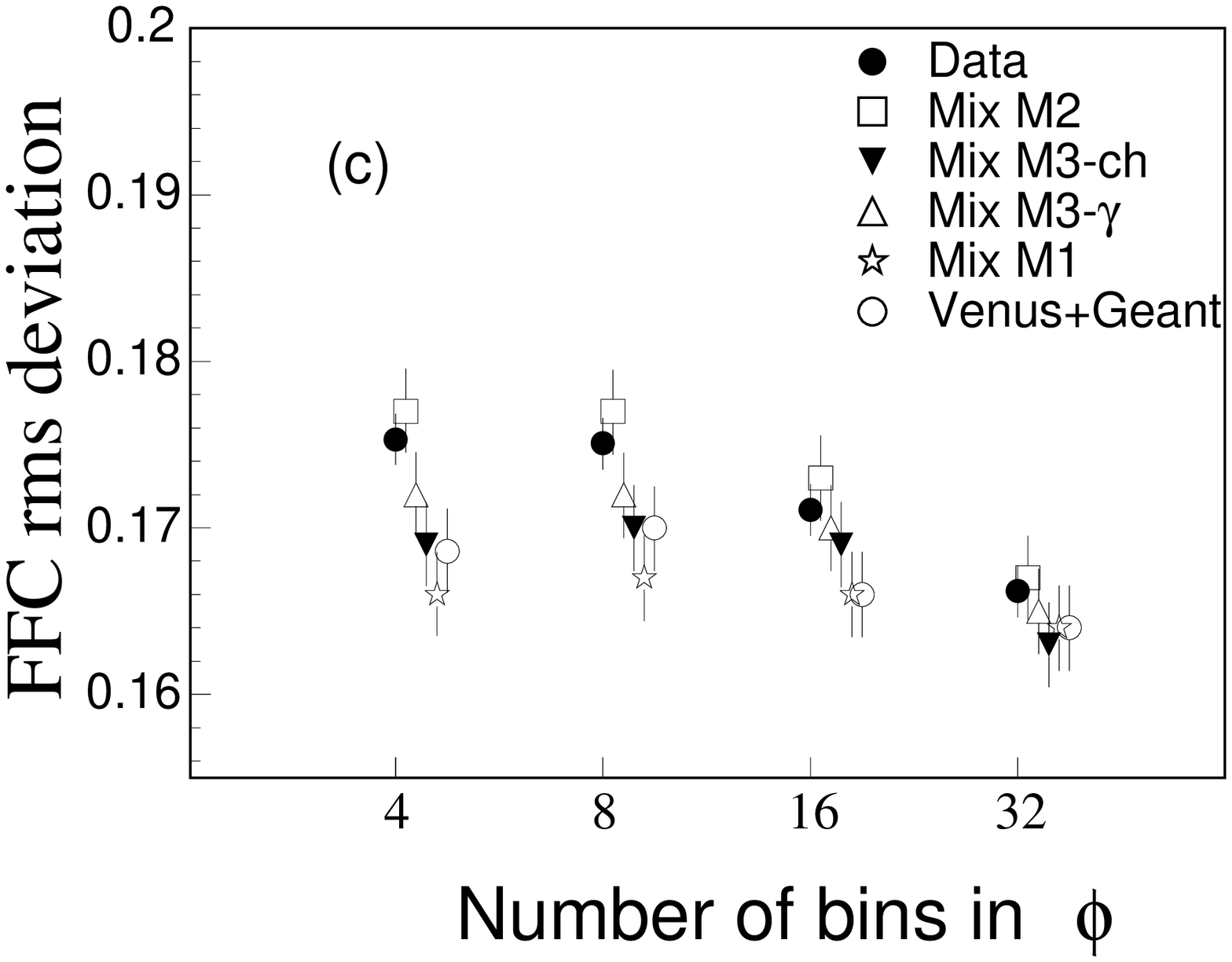}
\includegraphics[scale=0.4]{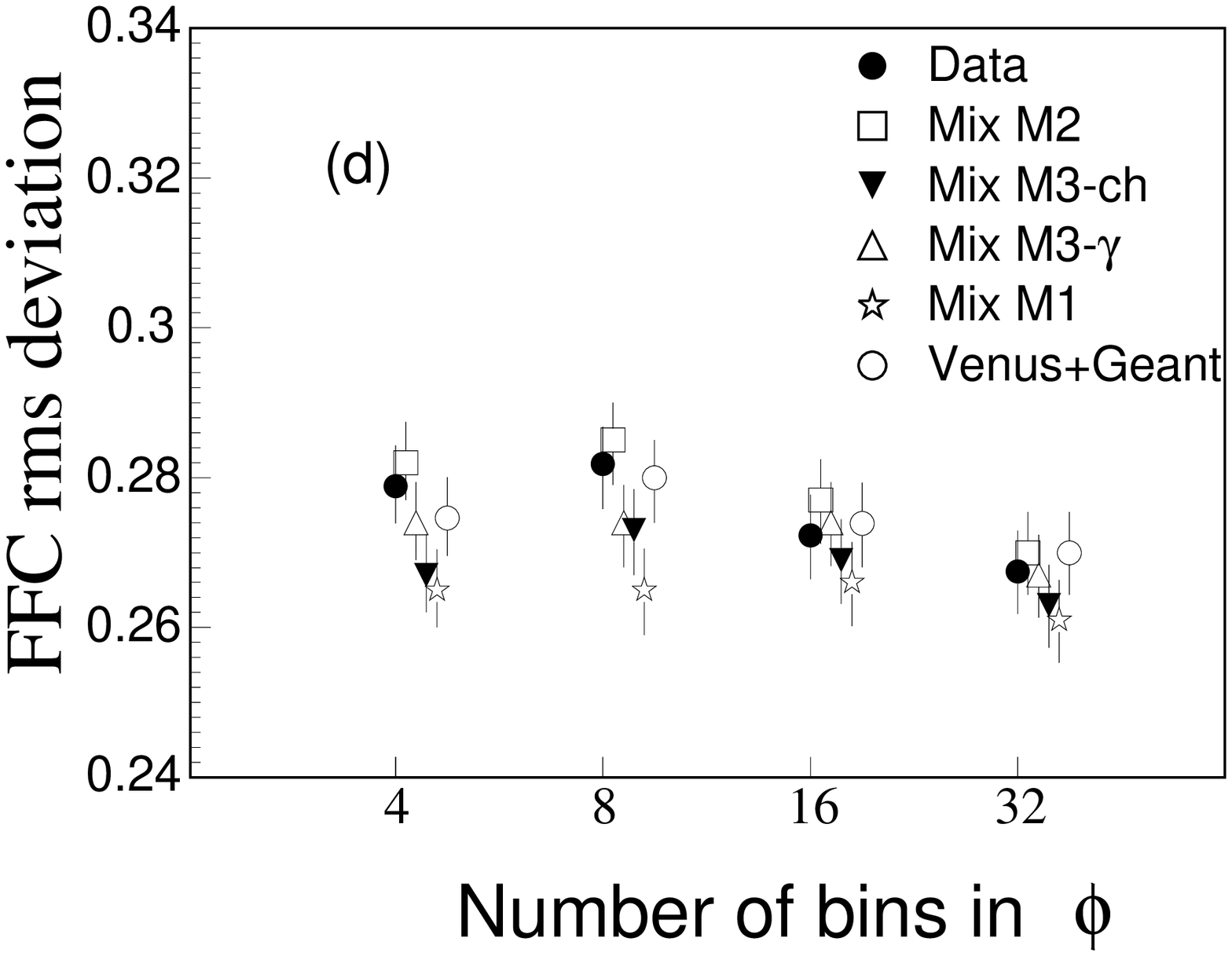}

\caption{ 
The RMS deviations of the FFC distributions for data,
various mixed events, and simulation for the four centrality bins
(a) centrality-1 ($0-5\%$), (b) centrality-2 ($5-10\%$), (c) 
centrality-3 ($15-30\%$), (d) centrality-4 ($45-55\%$).
}
\label{ffc_rms}
\end{center}
\end{figure*}

   The FFC distributions, at scales $j$=1 to 4, corresponding to 4 to
   32  bins
   in azimuthal angle, are shown in Fig.~\ref{ffc} for data,
   M1, and VENUS events for the four centrality classes. 
   The results for other types of mixed events
   are not shown for clarity of presentation.
   The widths of the FFC distributions are found to increase in
   going from the most central to most peripheral centrality class.
   The rms deviations of
   these FFC distributions are summarized in Fig.~\ref{ffc_rms}. 
   Similar to the case for the $S_{Z}$ distributions discussed above, 
   the rms deviations of the mixed events have been rescaled 
   by the percentage difference between the rms deviations of the 
   VENUS FFC distributions and those of the VENUS mixed
   events for each centrality class. 
   The statistical
   errors are small and are within the size of the symbols. 
   The bars represent statistical and systematic errors added in
   quadrature.

   The rms deviations of the FFC distributions for 
   the data, VENUS, and mixed events are found
   to be close to each other (within quoted errors) for the case of $32$ bins
   in $\phi$ for all of the four centrality classes. 
   The rms deviations for the  FFC distribution of 
   M2 mixed events are found to closely follow  those of the
   data for all centrality classes and all bins in $\phi$,
   while the rms deviations for the M3 mixed events lie
   between those of the data and M1 mixed events. 
   These results are consistent with those obtained from the 
   analysis of the $S_{Z}$
   distributions. These observations indicate the absence of
   event-by-event localized correlated fluctuations (DCC-like) between 
   $N_{\gamma-{\mathrm {like}}}$ and $N_{\mathrm {ch}}$. They also suggest
   the presence of localized fluctuations in both photons and charged
   particle multiplicities for intermediate bin sizes in azimuth.
   The rms values of the FFC distributions for VENUS events 
   are close to those of the M1 mixed events for centrality
   classes 1, 2, and 3. However they are slightly higher for
   the most peripheral centrality class (centrality-4).  

\subsection{Discussion}

   The results from the two independent methods of analysis
   are consistent and indicate the absence of
   event-by-event correlated DCC-like fluctuations in the photon and charged
   particle multiplicities.  However they do suggest the 
   presence of uncorrelated fluctuations in both the 
   photon and charged particle multiplicities for intermediate bin 
   sizes in $\phi$.
   The data has been compared to various kinds of mixed events and to simulated
   events which take into account many detector related effects.
   Still it is worthwhile to explore the extent of other possible experimental
   effects which might affect the observed rms 
   deviations of the $S_Z$ and FFC distributions. 
   As discussed extensively in Refs.~\cite{WA98-3,pmd_nim,WA98-9}, 
   care has been taken during the data taking and during the data processing 
   to closely monitor the performance of the PMD and SPMD.
   The detector uniformity of the PMD was studied in detail by using the
   minimum ionizing particle (MIP) signal from the data
   for all pads of the 22 boxes. The fluctuations of the pad-to-pad
   relative gains were approximately 10$\%$. The gain corrections were made 
   for each pad. Corrections for gain variations during
   the data taking period were made periodically for both the PMD and
   SPMD. This reduces the possibility of abrupt gain or threshold changes
   during the run period. Events with obvious detector readout effects, 
   such as missing or dead 
   regions, were carefully removed from the data sample.
   It should be recalled that most detector effects 
   are reflected in the mixed events.

   The effect of local fluctuations in the performance of the PMD 
   has been studied
   using simulated VENUS events. In one test, the gain
   of a group of pads corresponding to one or more PMD cameras
   was randomly varied. A 30$\%$ change of gain in one
   camera for all events resulted in an increase of 0.7$\%$
   in the rms deviations. Changing the gains of three cameras by
   30$\%$ for all events resulted in an increase of the rms
   deviations by 1.7$\%$.  Because the camera gains were closely monitored 
   on-line and during the processing of the data this is considered
   to be a highly unlikely scenario. Still, these changes are 
   within the quoted errors of Figs.~\ref{sz_rms}~and~ \ref{ffc_rms} 
   which indicates that local gain
   and threshold variations would not account for the observed differences
   between the data and the M1 mixed events.

   Several checks were performed to verify the quality of the data
   obtained with the SPMD. One of the differences between
   the previous analysis~\cite{WA98-3} and the present one is that,
   while in the previous case the analysis was performed
   using the total charged particle multiplicity of the detector
   deduced from the magnitude of the measured SPMD signals, here
   we simply used the total number of hit SPMD pads. The reason for
   using hit pads is that the correction in going from deposited charge to
   hits for each event and  small $\eta$--$\phi$ segments is
   non-trivial. 
   Also, the effects of two-track resolution
   and possible shifts of the beam position on the target during the
   spill were studied in detail. However
   all of these produced small effects which could not account for the
   differences observed between the data and M1 mixed events.

\subsection{Strength of localized fluctuations}

\begin{figure*}
\begin{center}
\includegraphics[scale=0.5]{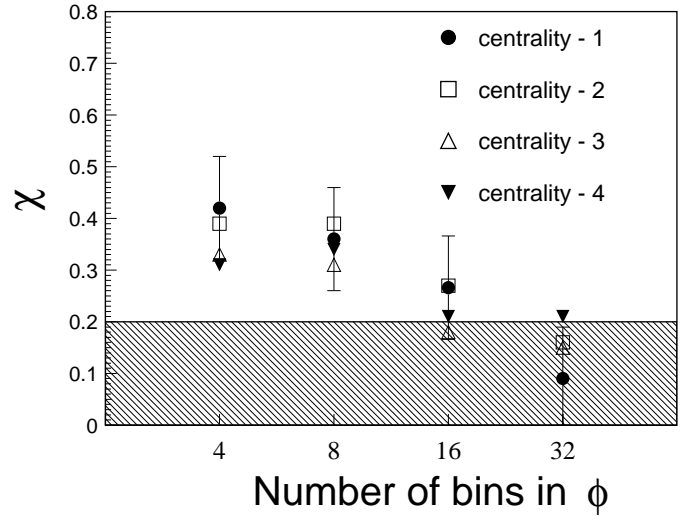}
\caption{ 
The fluctuation strength parameter  
for the four centrality classes. Centrality-1 corresponds to
the $5\%$ most central, 
centrality-2 corresponds to $5-10\%$, centrality-3 corresponds to 
$15-30\%$ and centrality-4 corresponds to $45-55\%$ of 
the minimum bias
cross section as determined by selection on the measured transverse energy
distribution. The error bars are
shown only on the centrality-1 selection for clarity of presentation. 
The errors are similar for the other centralities.
The shaded portion represents the limit above which a signal is detectable 
(see text for details).
}
\label{xi}
\end{center}
\end{figure*}

   In order to quantify the strength of the  
   $N_{\gamma-{\mathrm {like}}}$ and $N_{\mathrm {ch}}$ fluctuations
   for various bins in $\phi$ and for different centrality classes,
   we define a quantity $\chi$ as:

\begin{equation}
    \chi =  \frac{\sqrt{(s^2 - s_1^2)}}{s_1}
\label{chi_eqn}
\end{equation}
   where $s_1$ and $s$ correspond to the rms deviations
   of the FFC distributions of the M1 mixed events and
   real data, respectively. The results are shown in Fig.~\ref{xi} 
   as a function of the number of bins in $\phi$ for the four different 
   centrality
   classes. Qualitatively similar results are obtained when $\chi$ is 
   calculated using the rms deviations of the $S_{Z}$ distributions. 
   The shaded portion indicates the region of $\chi$ where $s$ is
   one $\sigma$ greater than the 
   rms deviation FFC distributions for M1 events, where $\sigma$ is 
   the total error on the M1 event rms deviation. It
   represents the limit above which a signal is detectable.
   Since $\chi$ is calculated from the rms deviations of the FFC 
   distributions for data and M1 mixed events it gives the combined  
   strength of localized fluctuations in both the photon and charged
   particle multiplicities. We do not present  $\chi$ values
   calculated using M3-type mixed events.  
   However, it is clear from the rms deviation figures 
   (Fig.~\ref{sz_rms} and \ref{ffc_rms})
   that both photons and charged particles contribute to the 
   observed fluctuations.
   The result shows that the strength of the fluctuations 
   decreases as the number of bins in $\phi$ increases, with a strength 
   which decreases to below detectable level (within the quoted errors) 
   for 16 and 32 bins. 
   There is an indication that the strength of the signal 
   decreases with decreasing centrality
   for 4 and 8 bins in azimuthal angle, 
   although the tendency is not very strong.

\subsection{Upper limit on DCC production}

It has been shown that the rms deviations of the 
$S_{Z}$ and FFC distributions for data are very close to those of the M2 
mixed events, within the quoted errors. If the DCC-like 
correlated fluctuations in
$N_{\gamma-{\mathrm {like}}}$ ${\mathrm {vs.}}$ $N_{\mathrm {ch}}$
were large, the rms deviations 
(Fig.~\ref{sz_rms} and Fig.~\ref{ffc_rms}) of data would have been larger
compared to those of the M2 mixed events. Since this is not the case, 
we may extract an
upper limit on the production of DCCs at the $90\%$ confidence
level following the standard procedure as discussed in Ref.~\cite{uplim}. 

The errors are assumed to have Gaussian distribution, although they are asymmetric.
The larger of the asymmetric errors is conservatively used 
for the limit calculation. 
The $90\%$ C.L upper limit contour has been calculated 
as $\chi$ + 1.28$e_{\chi}$, where $\chi$ is calculated 
using  Eq.(\ref{chi_eqn}). Here $s_1$ and $s$ correspond to the rms deviations
of the FFC (or $S_{Z}$) distributions for M2 mixed events and
real data, respectively, and $e_{\chi}$ is the error in $\chi$ from
the FFC 
(or $S_{Z}$) analysis. 
If the difference between the
rms deviation of the FFC distributions for M2 mixed events and real data is 
negative, we take the value of $\chi$ to be zero.
It may be mentioned 
that for the calculation of the upper limits we have 
assumed that the total difference
in the rms values of data and M2 mixed events is due to 
DCC-like fluctuations only. 

\begin{figure*}
\begin{center}
\includegraphics[scale=0.6]{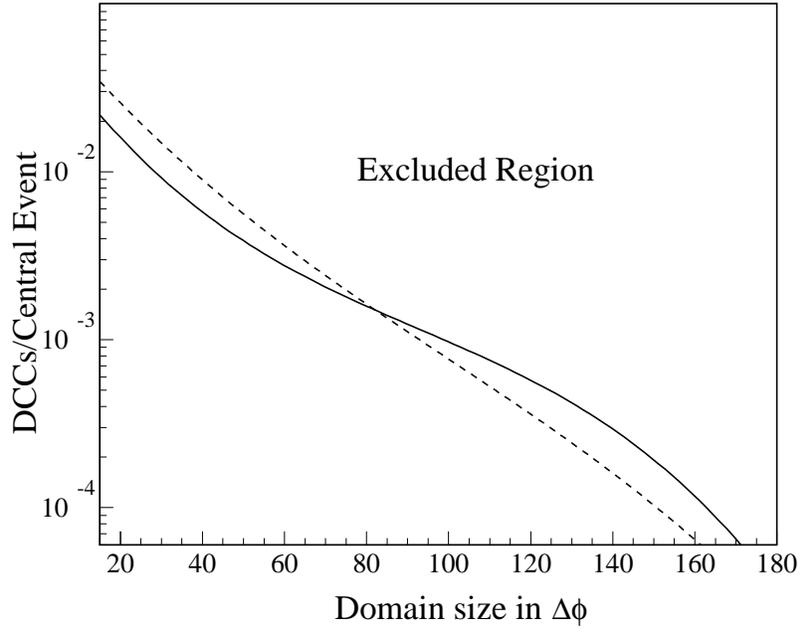}
\caption{The $90\%$ confidence level upper limit on DCC production for
central Pb+Pb collision at 158$\cdot$ A GeV/c, 
as a function of the DCC domain size in azimuthal angle
within the context of a simple DCC model and the measured photon and
charged particle multiplicities in the interval  $2.9< \eta < 3.75$. 
The solid line corresponds to data from the top $5\%$  and dashed line
to top $5-10\%$ of the minimum bias
cross section as determined by selection on the measured transverse energy
distribution. 
}
\label{upper_limit}
\end{center}
\end{figure*}

To relate the measured upper limit on the size of the fluctuations
to a limit on DCC domain size and frequency
of occurrence we proceed as follows: Within the context of the simple 
simulated DCC model described earlier, we obtain the rms deviations of 
the FFC  distributions with various domain sizes 
in azimuthal angle ($15^{0}$, $30^{0}$, $\cdots$ $180^{0}$) and for each 
domain size also for different frequency of occurrence of DCC ($0\%$ to $100\%$).
The M2 mixed events are then constructed for each of these sets of 
simulated events. For each set of DCC-type events of a given domain size
and frequency of occurrence, the value of $\chi$ is calculated using Eq.~\ref{chi_eqn},
from the difference in rms deviations of the FFC 
distribution of the DCC event ($s$) and its 
corresponding M2 mixed event distribution ($s1$). 
The upper limit is set at that 
value of frequency of occurrence for a fixed DCC domain size at which
the $\chi$ value from the DCC model matches with that of the 
$\chi$ + 1.28$e_{\chi}$ upper limit from the experimental data. 
This is used to set the upper
limit contour in terms of domain size and frequency of occurrence of 
the DCC. The  results for centrality classes 1 and 2 are shown in 
Fig.~\ref{upper_limit}. 
It may be mentioned that the 
upper limit contour set by a similar analysis of the rms values of
the  $S_{Z}$ distributions is very similar to that from the FFC analysis. 
Also, from  Fig.~\ref{xi} it is seen that the 
total fluctuation is similar or weaker for centrality classes 3 and 4,  hence 
the upper limit for these two classes would be similar or weaker than those 
shown in Fig.~\ref{upper_limit} for centrality classes 1 and 2.

\section{Summary}

     A detailed event-by-event analysis of the $\eta-\phi$ phase
    space distributions of the multiplicity of charged particles and photons in 
    Pb+Pb collisions at 158$\cdot A$~GeV has been carried out using 
    two different analysis methods for four different centrality classes.
    The results from the two
    analysis methods were found to be consistent with each other.
    The first analysis method studied the magnitude of the 
    $N_{\gamma-{\mathrm {like}}}$ versus $N_{\mathrm {ch}}$ multiplicity 
    fluctuations in decreasing phase space regions.  
    The second analysis employed the discrete wavelet transformation 
    technique to investigate the relative magnitude of the 
    $N_{\gamma-{\mathrm {like}}}$ versus $N_{\mathrm {ch}}$ fluctuations
    in adjacent phase space regions.  
    The results were compared to pure VENUS+GEANT simulation events and 
    to various types of mixed events to search for and identify the source of 
    non-statistical fluctuations. Both analysis methods indicated
    fluctuations beyond those observed in simulated and
    fully mixed events for $\phi$ intervals of greater than
    45$^\circ$ and which increased weakly in strength with increasing centrality.
    The additional fluctuations were found to be due to uncorrelated 
    fluctuations in both $N_{\gamma-{\mathrm {like}}}$ and $N_{\mathrm {ch}}$.
    No significant correlated fluctuations in $N_{\gamma-{\mathrm {like}}}$
    versus $N_{\mathrm {ch}}$, a likely signature of formation of 
    disoriented chiral condensates, were observed in all of the four centrality
    classes studied. Using the results 
    from the data, mixed events,
    and  within the limitations of a simple model of DCC formation, 
    an upper limit on DCC 
    production in 158$\cdot A$~GeV Pb+Pb collisions has been set.
   
\acknowledgments{

We wish to express our gratitude to the CERN accelerator division for the
excellent performance of the SPS accelerator complex. We acknowledge with
appreciation the effort of all engineers, technicians and support staff who
participated in the construction of this experiment.

This work was supported jointly by
the German BMBF and DFG,
the U.S. DOE,
the Swedish NFR and FRN,
the Dutch Stichting FOM,
the Polish KBN under Contract No. 621/E-78/SPUB-M/CERN/P-03/DZ211/,
the Grant Agency of the Czech Republic under contract No. 202/95/0217,
the Department of Atomic Energy,
the Department of Science and Technology,
the Council of Scientific and Industrial Research and
the University Grants
Commission of the Government of India,
the Indo-FRG Exchange Program,
the PPE division of CERN,
the Swiss National Fund,
the INTAS under Contract INTAS-97-0158,
ORISE,
Grant-in-Aid for Scientific Research
(Specially Promoted Research \& International Scientific Research)
of the Ministry of Education, Science and Culture,
the University of Tsukuba Special Research Projects, and
the JSPS Research Fellowships for Young Scientists.
ORNL is managed by UT-Battelle, LLC, for the U.S. Department of Energy
under contract DE-AC05-00OR22725.
The MIT group has been supported by the US Dept. of Energy under the
cooperative agreement DE-FC02-94ER40818.

}

%\onecolumn 




\end{document}